\documentclass[epj]{svjour}

\usepackage{graphics}
\usepackage{epsfig}

\begin{document}

\title{Attraction of like-charged macroions in the strong-coupling 
limit}

\author{Ali Naji
      \thanks{Present address: Physics Department, Technical University of Munich, 
       85748 Garching, Germany; \email{naji@ph.tum.de}}
       \and 
       Roland R. Netz
       \thanks{\email{netz@ph.tum.de}}
       }

\institute{Sektion Physik, Ludwig-Maximilians-Universit\"at, 
          Theresienstr. 37, D-80333 M\"unchen, Germany.   
          \\
          Max-Planck-Institut f\"ur Kolloid- und 
          Grenzfl{\"a}chenforschung, Am M\"uhlenberg, D-14476 Golm, Germany.
          }

\date{Ref.: Revised version published at Eur. Phys. J. E {\bf 13}, 43 (2004).}

\abstract{Like-charged macroions attract each other as a result 
of strong electrostatic correlations
in the presence of multivalent counterions or at low temperatures. 
We investigate the effective electrostatic 
interaction between i) two like-charged rods and ii) two like-charged
spheres using the recently introduced strong-coupling theory,
which becomes asymptotically exact in the limit of large coupling parameter
({\em i.e.} for large counterion valency, low temperature, or high surface charge 
density on macroions). In contrast to previous applications 
of the strong-coupling theory, we deal with
curved surfaces and an additional parameter, referred to 
as Manning parameter, is introduced, which measures the ratio between the
radius of curvature of macroions to the Gouy-Chapman length. 
This parameter, together with the size of the
confining box enclosing the two macroions and their neutralizing counterions, 
controls the counterion-condensation process that directly affects the effective 
interactions.
For sufficiently large Manning parameters (weakly-curved surfaces), we find 
a strong long-ranged attraction between two macroions that form 
a closely-packed bound state with small 
surface-to-surface separation of the order of the counterion diameter
in agreement with recent simulations results. 
For small Manning parameters (highly-curved surfaces), 
on the other hand, the equilibrium separation increases and the
macroions unbind from each other as the confinement volume increases to infinity. 
This occurs via a continuous universal unbinding transition 
for two charged rods at a threshold Manning parameter of
$\xi_c = 2/3$, while the transition is strongly discontinuous for spheres 
because of a pronounced potential barrier at intermediate distances. 
Unlike the cylindrical case, the attractive forces between spheres 
disappear  for increasing confinement volume due to the
complete de-condensation of counterions. 
Scaling arguments suggest that for moderate values of coupling parameter, 
strong-coupling predictions remain valid for sufficiently small
surface-to-surface separations. 
\PACS{{87.15.-v}{Biomolecules: structure and physical properties} \and
      {82.70.Dd}{Colloids} \and
      {87.15.Nn}{Properties of solutions; aggregation and crystallization 
of macromolecules}}
}

\maketitle




\section{Introduction}
\label{sec:intro}

Electrostatic interactions play a prominent role in many  
soft-matter and biological systems, since a huge class of macromolecules
are water-soluble, and thus bear electric charges in
aqueous solutions. These macroions may be 
polyelectrolytes such as DNA, or charged colloidal particles, that are
surrounded by their neutralizing (oppositely-charged) 
counterions. In recent years, mounting evidence 
from both experiments 
\cite{Bloom,Delsanti,Della,Strey98,Tang,Tang03}    
and numerical simulations 
\cite{Guld84,Svensson,Guld86,Wood88,Valleau,Lyub95,Gron97,Gron98,Wu,AllahyarovPRL,Stevens99,LinsePRL,Lise00,Hribar,Messina00,Allahyarov00,Deserno03,Arnold03,Arnold-Naji} 
showed that like-charged macroions can attract each other 
via effective forces of electrostatic origin, notorious examples of which 
are formation of dense packages of DNA molecules (DNA condensates) \cite{Bloom} 
and  large aggregates of charged colloidal particles \cite{LinsePRL,Lise00,Hribar}.
These observations indicate the appearance of like-charge attraction in 
strongly-charged systems, {\em i.e.} when multivalent counterions are present,
macroions are highly charged, or the strength of electrostatic 
interactions is enhanced by maintaining the system at low temperatures or
in a medium of low dielectric constant 
\cite{Bloom,Guld84,Svensson,Guld86,Wood88,Valleau,Lyub95,Gron97,Gron98,Wu,AllahyarovPRL,Stevens99,LinsePRL,Lise00,Hribar,Messina00,Allahyarov00,Deserno03,Arnold03,Arnold-Naji}.

On the other hand, there has been a great deal of theoretical studies
aimed at revealing possible underlying mechanisms, which can lead to such 
counter-intuitive interactions 
\cite{Guld84,Svensson,Guld86,Wood88,Valleau,Lyub95,Gron97,Gron98,Wu,AllahyarovPRL,Stevens99,LinsePRL,Lise00,Hribar,Messina00,Allahyarov00,Deserno03,Arnold03,Arnold-Naji,Kjellander84,kjellander92,Oosawa,Attard-etal,Pordgornik90,Barrat,Pincus98,Podgornik98,Ha,Safran99,Kardar99,Netz-orland,Ha01,Lau02,Stevens90,Diehl99,Rouzina96,Korny,Arenzon99,Shklovs99,Lau01,Netz01,AndrePRL,Manning97,Rouzina98,Golestan99,Levin99,Belloni95}.
Historically, electrostatic interactions in charged systems have been 
studied by means of mean-field approximations, such as Poisson-Boltzmann (PB) 
(or its linearized version, Debye-H\"uckel) theory \cite{VO,DH}. 
Despite the fact that in many instances, 
the PB theory provides a successful
approximation, it has been shown rigorously \cite{Neu,Sader,Trizac} that 
such mean-field theories can only predict repulsion between
like-charged objects.  Recent 
investigations have, however, elucidated the important role 
of electrostatic correlations, neglected
in the mean-field approximation, toward understanding the mechanism of like-charge 
attraction. Several proposals have been put forward in order to 
incorporate ionic correlations. 
These attempts include integral-equation approach 
\cite{Kjellander84,kjellander92}, 
perturbative improvement of the PB theory 
\cite{Oosawa,Attard-etal,Pordgornik90,Barrat,Pincus98,Podgornik98,Ha,Safran99,Kardar99,Netz-orland,Ha01,Lau02},
local density-functional theory \cite{Stevens90,Diehl99}, 
structural-correlation approach \cite{Rouzina96,Korny,Arenzon99,Shklovs99}, 
harmonic-plasmon approximation \cite{Lau01},
and the strong-coupling theory \cite{Netz01,AndrePRL} 
(that will be used in this paper), which generally compare well
with simulation results and all exhibit attraction. 
Nonetheless, they are different in technical aspects, 
and have their own range of rigor and applicability.

In general, the
importance of electrostatic correlations can be quantified by means of the
{\em coupling parameter} $\Xi=2\pi q^3 \ell_B^2 \sigma_s$  \cite{Netz01}, which
depends on the charge valency of counterions  $q$,  
surface charge density of macroions $\sigma_s$, and the Bjerrum length  
$\ell_B=e^2/(4\pi \varepsilon\varepsilon_0 k_B T)$
(associated with a medium of dielectric constant 
$\varepsilon$ and at temperature $T$). The PB theory 
is asymptotically obtained in the limit of vanishingly small coupling 
parameter $\Xi\rightarrow 0$
\cite{Netz-orland}, while, non-mean-field features  emerge in the opposite limit 
of large coupling parameter, $\Xi\gg1$, and are 
accompanied by strong accumulation of  counterions 
in the proximity of charged objects, where counterions may also form
strongly-correlated structures, such as a quasi-two-dimensional liquid
or possibly a Wigner crystal at low temperatures 
\cite{AllahyarovPRL,Lise00,Rouzina96,Shklovs99,Lau01,AndrePRL}.

For rod-like and spherical macroions,  
effective interactions are also influenced by the 
entropy-driven counterion-condensation process, which
basically arises as a result of the specific interplay between energetic 
and entropic contributions
in these geometries. Clearly, a dominant attraction between like-charged 
macroions necessitates
a sufficiently large number of counterions being condensed in their close 
vicinity. The counterion-condensation process at rod-like macroions 
is controlled by the Manning parameter \cite{Manning69} defined here as 
$\xi=q \ell_B \tau$, where $\tau$ stands for the single-rod linear charge 
density. For small Manning parameter, counterions
de-condense leading to a bare electrostatic 
repulsion between charged rods regardless of the coupling parameter. While
for sufficiently large values of Manning parameter, an 
attractive force may prevail as  a certain fraction of counterions
remains bound to the rods even in the absence of confining boundaries. 
An interesting problem is, therefore, to  examine the 
regime of Manning parameters, where an effective attraction emerges. 
This has been addressed in a number of previous works for a pair of 
like-charged rods. It was argued by Ray and Manning \cite{Manning97}
that attraction sets in, when the Manning parameter becomes 
larger than the threshold of counterion condensation
in the two-rod system, {\em i.e.} $\xi=1/2$. While, a 
different attraction range of $\xi>2$ was proposed by the 
counterion-condensation theory of Arenzon {\em et al.} \cite{Arenzon99}. 
Numerical simulations, on the other 
hand, reveal attraction between two like-charged rods
in a wide range of Manning parameters 
including $\xi\approx 1$ (for moderate coupling parameters) 
\cite{Gron97,Deserno03,Arnold03,Arnold-Naji}, though 
have not yet specified the threshold value.

For a system of charged spheres, attraction is induced 
only when the system is confined in a box, because
the energetic binding of counterions to charged spheres 
is weaker than the entropic gain resulting in 
complete de-condensation of counterions in the 
absence of confining boundaries. The influence of the confinement 
volume on the effective interaction
between like-charged spheres 
was addressed in the simulations by Gr{\o}nbech-Jensen {\em et al.} 
\cite{Gron98}. It was shown that the pair potential of mean force
develops a local minimum at 
small surface-to-surface separation, which corresponds to
a closely-packed bound state of spheres maintained by 
a strong short-ranged attraction, as indicated independently by other 
simulations \cite{Wu,AllahyarovPRL}. This attraction regime is 
separated from a repulsion regime at large separations
by a pronounced potential barrier \cite{Gron98}. 
Upon increasing the confinement volume, the strength of attraction  
and the barrier height decrease leading to pure repulsion between spheres
in a very large confining box. Other simulations on many-sphere systems
\cite{LinsePRL,Lise00,Hribar} have also displayed formation of large aggregates 
of highly-charged spheres, which indicate a phase separation in the thermodynamic limit  
\cite{Gron98,LinsePRL,Lise00}.  As discussed by Linse {\em et al.} \cite{Lise00},
the equilibrium surface-to-surface separation of the attracting spheres in 
the compact state appears to be of the order of the counterion diameter.

In this paper, we investigate the effective interaction
between a pair of like-charged rods and a pair of like-charged 
spheres using the strong-coupling (SC) theory introduced in previous works 
\cite{Netz01}. In contrast to previous applications of the strong-coupling theory
\cite{Netz01,AndrePRL}, we deal with curved surfaces and thus, 
an additional parameter is introduced, which measures the ratio between 
the radius of curvature of macroions to the associated Gouy-Chapman length.
For charged rods, this parameter 
is equivalent with the so-called Manning parameter (introduced above), and thus, 
we shall refer to this additional parameter as {\em Manning parameter} 
for both charged rods and charged spheres. 
We shall identify the regime of Manning parameters, where attraction 
is predicted between macroions with or without any confinement,
and examine the equilibrium separation of interacting macroions
for varying Manning parameter and confinement volume.

The SC theory is obtained by 
a systematic virial expansion in the limit of large coupling
parameter, $\Xi\rightarrow \infty$, which is
complementary to the mean-field regime (Section \ref{sec:SCgeneral}). 
This, in particular,  covers the low-temperature regime 
(but still above the crystallization temperature  \cite{AndrePRL}). 
Nonetheless, the theory involves finite entropic contributions, thus 
finite-temperature effects from counterions that appear at leading order
in a simple one-particle form. This 
allows for reproducing the de-condensation process of counterions 
at low Manning parameters and therefore, leads to
a consistent  picture in the whole range of Manning parameters.
For sufficiently large Manning parameters, 
when counterions effectively condense around macroions, 
we find a long-ranged attraction, which 
originates mainly from the energetic attraction mediated 
by counterions intervening in a narrow region 
between macroions. The effective force is dominated by such 
an attraction for increasing Manning parameter (or decreasing temperature)
and tends to a temperature-independent value, which scales with the 
macroion separation, $D$, as $\sim 1/D$ for two charged rods, 
and as $\sim 1/D^2$ for two charged spheres. 
(For two charged plates, the effective force is obtained to be independent
of $D$ \cite{Netz01,AndrePRL}.) In this regime, two macroions are predicted to form 
a closely-packed bound state with a surface-to-surface
distance of the order of the counterion diameter in qualitative 
agreement with numerical findings 
\cite{Gron98,Wu,LinsePRL,Lise00,Deserno03,Arnold03}.  
Such an agreement is established, when the excluded-volume repulsion 
between counterions is negligible compared to the  electrostatic interactions.
For point-like counterions, the equilibrium surface-to-surface distance is
predicted to be of the order of the Gouy-Chapman length --see 
Eqs. (\ref{eq:Delta_rods}) and (\ref{eq:Delta_sph}) below-- and 
a quantitative agreement is obtained between SC 
predictions and simulation results as reported elsewhere \cite{Arnold-Naji}.

We shall also study the influence of the confinement due to
an outer box on the effective interaction and on the onset of 
attraction. It will be shown that by decreasing the Manning parameter or 
increasing the size of the confinement, macroions undergo an unbinding
transition. For two charged rods, 
the minimal Manning parameter to obtain attraction (the threshold Manning parameter) 
tends to the universal value of $\xi_c=2/3$ for increasing box size, which 
is somewhat {\em larger} than the corresponding onset 
of counterion condensation $\xi=1/2$.
The unbinding transition in this system occurs continuously
and is characterized by a power-law behavior. 
In contrast, charged spheres display a discontinuous transition 
and a potential barrier in the effective free energy in qualitative 
agreement with simulation results \cite{Gron98}. It turns out that 
the attraction threshold for two like-charged spheres 
monotonically increases with the size of the confining
box yielding a pure repulsion between unconfined spheres, but its 
dependence on the box size 
is logarithmically weak.  Such a weak 
dependence of the onset of attraction on the box size may  explain
the stability of aggregates of highly-charged spheres 
in large confinements that has been observed in recent
simulations \cite{LinsePRL,Lise00,Hribar}. 
The range of applicability of the present theory to systems 
with finite coupling parameter will be examined by means of scaling 
arguments.




\section{Strong-Coupling Theory: General formalism}
\label{sec:SCgeneral}
 
Consider a system of macroions with charge distribution 
$\sigma({\mathbf r})$ (in units of the elementary charge
$e$), and a  number of $N$ oppositely-charged counterions 
of diameter $\sigma_c$ and charge valency $q$. 
All charges interact via Coulombic interaction 
$v({\mathbf r})=1/|{\mathbf r}|$, and
the electroneutrality condition is assumed to hold globally
in the system. Hence, one always has
\begin{equation} 
Nq=\int d{\mathbf r}\sigma({\mathbf r}).
\label{eq:elecneut}
\end{equation}
Note that $q$ and $\sigma({\mathbf r})$ are defined to be positive.
We shall suppose that macroions
are fixed. (In the forthcoming calculations in Sections \ref{sec:rods} and \ref{sec:spheres}, 
we shall consider excluded-volume interactions of hard-core nature between
particles--see Section \ref{subsec:SCfree}.)
 Assuming a surface charge density of 
$\sigma_s$ for macroions, the Gouy-Chapman
length, $\mu$, is defined as 
\begin{equation}
  \mu=\frac{1}{2\pi q\ell_B\sigma_s},
\label{eq:GC}
\end{equation} 
which sets a
characteristic length scale for the considered system. Here, 
$\ell_B=e^2/(4\pi \varepsilon\varepsilon_0 k_BT)$ is the
Bjerrum length, the distance at which two elementary charges 
interact with thermal energy $k_B T$ in a medium of dielectric constant 
$\varepsilon$.

It is convenient to construct a dimensionless formalism, which may be 
achieved by rescaling spatial coordinates, say ${\mathbf r}$, with the 
Gouy-Chapman length as ${\tilde \mathbf r}={\mathbf r}/\mu$. 
Other parameters are rescaled accordingly, for instance, 
\begin{equation}
  {\tilde \sigma}({\tilde {\mathbf r}})=\mu\sigma({\mathbf r})/\sigma_s.
\label{eq:sigma}
\end{equation}
In the rescaled units and for point-like counterions ($\sigma_c=0$), 
the grand-canonical partition function of the system, 
${\mathcal Z}_\Lambda$, is mapped to
a one-parametric field theory (apart from rescaled 
geometrical parameters such as the typical macroion size), 
and may be expressed as a functional integral over a fluctuating field 
$\phi({\tilde {\mathbf r}})$ \cite{Netz01}
\begin{equation}
  {\mathcal Z}_\Lambda=
        \int\,\frac{{\mathcal D}\phi}{{\mathcal Z}_v}
                \exp\{-{\tilde {\mathcal H}}[\phi]/\Xi\},
\label{eq:Z}                 
\end{equation}
where 
\begin{equation}
  \Xi=2\pi q^3 \ell_B^2 \sigma_s
\label{eq:Xi}
\end{equation}
is the coupling parameter,
${\mathcal Z}_v=\sqrt{{\mathrm{Det}}\, v}$ contains self-energy
contributions and the action reads
\begin{equation}
  {\tilde {\mathcal H}}[\phi]=
                   \frac{1}{2\pi}
                    \int d{\tilde {\mathbf r}}
                        \left[
                           \frac{1}{4}(\nabla \phi({\tilde {\mathbf r}}))^2
                              -\imath \phi({\tilde {\mathbf r}})
                                      {\tilde \sigma}({\tilde {\mathbf r}})
                              -\Lambda{\tilde \Omega}({\tilde {\mathbf r}})
                                    e^{-\imath \phi({\tilde {\mathbf r}})}
                        \right].                                             
\label{eq:action}
\end{equation}
Here, $\Lambda$ is the rescaled fugacity, and 
the  function ${\tilde \Omega}({\tilde {\mathbf r}})$ takes
geometrical constraints into consideration, {\em e.g.} restricts the 
positions of mobile counterions to an appropriate region in space.
The electroneutrality condition, Eq. (\ref{eq:elecneut}), written
in rescaled units, relates the coupling parameter 
to the average number of counterions through
\begin{equation}
  {\tilde Q}\equiv
            \int d{\tilde {\mathbf r}}{\tilde \sigma}({\tilde {\mathbf r}})
            =2\pi\Xi N,
\label{eq:Q}
\end{equation} 
where ${\tilde Q}$ is simply the rescaled area of macroions covered
by electric charges.

In general, statistical quantities such as counterions density profile,
osmotic pressures and effective forces can be 
obtained from Eq. (\ref{eq:Z}). 
But, as exact calculations based on the partition function (\ref{eq:Z}) 
are difficult, one may consider limiting cases, 
where approximate methods are applicable. For instance, at
small coupling parameters  $\Xi\rightarrow 0$, a
saddle-point approximation may be applied, since  
the functional integral in this limit is dominated by
the saddle point of the action, Eq. (\ref{eq:action}).
This approximation yields the so-called mean-field  
Poisson-Boltzmann (PB) theory, which may be then extended to  
finite couplings by means of a systematic loop expansion  around the 
saddle-point solution \cite{Netz-orland}. However,
the loop expansion around PB theory was shown to be a weakly 
convergent series and can not improve the PB results to be applicable 
for systems of large coupling parameter \cite{AndrePRL}.

At large coupling parameters $\Xi\gg 1$, a series expansion can 
be obtained for the partition function (\ref{eq:Z}) 
in powers of the rescaled fugacity, $\Lambda/\Xi$, as
\cite{Netz01}
\begin{eqnarray}
  {\mathcal Z}_\Lambda=
                {\mathcal Z}_0
                \sum_{j=0}^{\infty}
                    \frac{1}{j!}\left[\frac{\Lambda}{2\pi\Xi}\right]^j
                \prod_{k=1}^{j}
                          \left[\int\,d{\tilde {\mathbf r}}_k
                             {\tilde \Omega}({\tilde {\mathbf r}}_k)\right]
         \nonumber\\
                \times\exp\{
                    -\Xi\sum_{n<m}^j v({\tilde {\mathbf r}}_n-
                                        {\tilde {\mathbf r}}_m)
                    -\sum_{i=1}^j {\tilde u}({\tilde {\mathbf r}}_i)
                  \},          
\label{eq:ZSC}
\end{eqnarray}  
which is nothing but a virial expansion with respect to the 
counterionic degrees of freedom. The zeroth-order term of the expansion,
${\mathcal Z}_0$, represents the partition function of the system, when
all counterions are taken out and only the fixed 
charge distribution is present,
\begin{equation}
  {\mathcal Z}_0=e^{-{\tilde U}_0/\pi\Xi},
\label{eq:Z0}
\end{equation}
where
\begin{equation}
  {\tilde U}_0=
           \frac{1}{8\pi}\int\,d{\tilde {\mathbf r}}d{\tilde {\mathbf r}}'
             {\tilde \sigma}({\tilde {\mathbf r}})
                v({\tilde {\mathbf r}}-{\tilde {\mathbf r}}')
                  {\tilde \sigma}({\tilde {\mathbf r}}')
          -\frac{{\tilde Q}}{4\pi}\int\,d{\tilde {\mathbf r}}'
                v({\tilde {\mathbf r}}'-{\tilde {\mathbf r}}_0)
                  {\tilde \sigma}({\tilde {\mathbf r}}')
\label{eq:U0}
\end{equation}
is the rescaled zero-particle interaction energy. (Note that 
${\tilde {\mathbf r}}_0$ is an arbitrary point chosen to 
fix the reference configuration with respect to which
the interaction energy is calculated.)

The first-order term in Eq. (\ref{eq:ZSC}) 
is the partition function of the system in the presence of 
a single counterion interacting with the fixed charge 
distribution via the one-particle interaction
\begin{equation}
  {\tilde u}({\tilde {\mathbf r}})=
        -\frac{1}{2\pi}\int\,d{\tilde {\mathbf r}}'                
                  [v({\tilde {\mathbf r}}-{\tilde {\mathbf r}}')
                 -v({\tilde {\mathbf r}}_0-{\tilde {\mathbf r}}')]
                {\tilde \sigma}({\tilde {\mathbf r}}').         
\label{eq:u}
\end{equation}
Higher-order terms in Eq. (\ref{eq:ZSC})
involve the two-particle interaction
$v({\tilde {\mathbf r}}-{\tilde {\mathbf r}}')$, that emerges in the
non-perturbative form of $\exp(-\Xi v)$.

The canonical strong-coupling (SC) theory is obtained as an asymptotic 
theory from the above expansion, Eq. (\ref{eq:ZSC}),
in the limit of $\Xi\rightarrow \infty$. 
As we shall see below (Section \ref{subsec:SCfree}), the leading-order term in 
the canonical free energy contains counterionic contributions only up 
to the one-particle terms. 
This reflects the fact that for large $\Xi$, counterion-macroion 
correlations dominate over counterion-counterion correlations.
Higher-order corrections for finite couplings
have been studied analytically for systems composed of 
planar charged walls in Refs. \cite{Netz01,AndrePRL} and will not be 
considered in the present study.

Clearly, the virial expansion, Eq. (\ref{eq:ZSC}), could have been 
obtained directly from the original partition function 
of the system. Also note that although the field-theoretic representation of the 
grand-canonical partition function, Eq. (\ref{eq:Z}), was obtained for
point-like counterions, the virial expansion, Eq. (\ref{eq:ZSC}), is quite general 
and can systematically include finite counterion size ($\sigma_c\neq 0$) via
excluded-volume interactions between particles (see Section \ref{subsec:SCfree}).
Taking the detour over the field-theoretic
formulation based on Eqs. (\ref{eq:Z}) and (\ref{eq:action}), however, 
demonstrates that loop expansion 
and virial expansion are indeed the two asymptotic limits of the same 
problem: While the PB theory and loop expansion 
become asymptotically exact in the limit of weak coupling
$\Xi\rightarrow 0$, the SC 
theory and the virial
expansion produce asymptotically exact results in the complementary limit 
of large coupling $\Xi\rightarrow \infty$.
One may notice that higher-order terms in Eq. (\ref{eq:ZSC})
contain integrations over the so-called Mayer function $\exp(-\Xi v)-1$, 
that are weighted by other
factors coming from the interaction of counterions with 
macroions. Such integrations are known to be divergent 
for bulk systems such as electrolytes, where  
the virial expansion fails to produce 
a convergent low-density expansion for the equation of state.  
In contrast, here counterions 
are considered at charged macroscopic objects. It has been 
shown explicitly for charged planar walls that in the limit of 
$\Xi\rightarrow \infty$, all orders of the virial expansion produce
finite contributions to the density profile of counterions and the effective 
force between walls \cite{Netz01}.


\subsection{The strong-coupling free energy}
\label{subsec:SCfree}

To calculate the strong-coupling free energy, we 
start from the grand-canonical free
energy (in units of $k_BT$),
\begin{equation}
  {\mathcal Q}_\Lambda=-\ln {\mathcal Z}_\Lambda, 
\end{equation}
where $Z_\Lambda$ is given by Eq. (\ref{eq:ZSC}). 
The Legendre transformation 
\begin{equation}
  {\mathcal F}_N=N\ln \Lambda+{\mathcal Q}_\Lambda,
\label{eq:legendre}
\end{equation}
provides us with the canonical free energy 
 ${\mathcal F}_N$.
The rescaled fugacity, $\Lambda$, in Eq. (\ref{eq:legendre})
is calculated from
\begin{equation}
   N=\Lambda\frac{\partial \ln{\mathcal Z}_\Lambda}{\partial \Lambda},
\label{eq:NLambda}
\end{equation}
in terms of  $\Xi$ and other rescaled geometrical factors. (Note that
$N$ can be eliminated using Eq. (\ref{eq:Q}).) In general, 
we may propose the following expression for 
$\Lambda$ in the large coupling limit $\Xi\gg 1$,
\begin{equation}
  \Lambda=\Lambda_0+\frac{\Lambda_1}{\Xi}+\frac{\Lambda_2}{\Xi^2}+\ldots,
\label{eq:lambdaexpand}
\end{equation}
where $\Lambda_0, \Lambda_1,\ldots$ are 
determined from Eq. (\ref{eq:NLambda}) using Eqs. (\ref{eq:Q}), 
(\ref{eq:ZSC})-(\ref{eq:u}) \cite{Note1}.  
It is easy to verify that, for instance,
\begin{eqnarray}
   \Lambda_0 &=& \frac{\tilde Q}
                        {
                          \int\,d{\tilde {\mathbf r}}
                           {\tilde \Omega}({\tilde {\mathbf r}})
                            e^{-{\tilde u}({\tilde {\mathbf r}})}
                        },
    \label{eq:lambda0}
    \\
    \Lambda_1 &=& \frac{
                        {\tilde Q}^2 
                          \int\,d{\tilde {\mathbf r}}d{\tilde {\mathbf r}}'
                            {\tilde \Omega}({\tilde {\mathbf r}})
                            {\tilde \Omega}({\tilde {\mathbf r}}')
                             e^{-{\tilde u}({\tilde {\mathbf r}})-
                                  {\tilde u}({\tilde {\mathbf r}}')}
                       [1-e^{-\Xi v({\tilde {\mathbf r}}-{\tilde {\mathbf r}}')}]
                       }
                       {
                        2\pi\left[
                        \int\,d{\tilde {\mathbf r}}
                           {\tilde \Omega}({\tilde {\mathbf r}})
                            e^{-{\tilde u}({\tilde {\mathbf r}})}\right]^3
                        }.
\end{eqnarray} 
Inserting this into Eq. (\ref{eq:legendre}), 
we find the canonical free
energy ${\mathcal F}_N$ (in units of  $k_BT$ and up 
to an irrelevant additive constant), 
which also admits a large-coupling expression
\begin{equation}
  {\mathcal F}_N=  \frac{{\mathcal F}_1}
                                       {\Xi}
                                   +\frac{{\mathcal F}_2}
                                       {\Xi^2}
                                   +\ldots,
\label{eq:FEexpansion} 
\end{equation}
where the coefficient of the leading-order term is
\begin{equation}
   {\mathcal F}_1 = \frac{{\tilde U}_0}{\pi}
                           -\frac{{\tilde Q}}{2\pi}
                                \ln\int\,d{\tilde {\mathbf r}} 
                                    {\tilde \Omega}({\tilde {\mathbf r}})
                                      e^{-{\tilde u}({\tilde {\mathbf r}})}
                           +C_0,
\label{eq:FESC}
\end{equation}
in which $C_0=(\tilde Q/2\pi)\ln {\tilde Q}-\tilde Q/2\pi$ is a constant, and
the geometry function ${\tilde \Omega}$, as introduced before,
specifies the accessible volume for counterions.

The free energy coefficient ${\mathcal F}_1$, which we may refer to as
the {\em rescaled  SC free energy} of the system, 
yields the actual {\em strong-coupling free energy}, 
${\mathcal F}^{SC}_N$, that is 
\begin{equation}
   {\mathcal F}^{SC}_N=\frac{{\mathcal F}_1}{\Xi},
\label{eq:FSCF1}
\end{equation}
which generates the leading (non-vanishing) contribution to the effective forces
between macroions \cite{Note2}. 
This will be used in the following Sections to investigate the effective
interaction between two charged rods and two charged spheres in the
strong-coupling limit.  The SC free energy has  
a non-trivial temperature dependence that will be discussed briefly
in the Appendix for some asymptotic cases.

For realistic systems with {\em finite} coupling parameter, 
higher-order corrections to the SC free energy become increasingly 
important as in this situation, counterion-counterion correlations 
become comparable to the counterion-macroion correlations 
accounted for by the asymptotic term. 
Thus the range of applicability of SC predictions at moderate to large couplings 
should be examined by calculating higher-order corrections. 
These calculations are, however, difficult to perform
analytically for the system of two charged
rods and two charged spheres  considered in this paper. Thus, we shall use
qualitative schemes to identify the regime of validity of SC results 
based on the fact that electrostatic correlations induced by counterions
between apposing macroions are dominant, when
the typical lateral distance between counterions becomes larger than 
or comparable to the macroions surface-to-surface separation.
This criterion, which was first introduced by Rouzina and Bloomfield 
\cite{Rouzina96}, has been derived by calculating higher-order terms in 
Eq. (\ref{eq:ZSC}) for planar charged walls 
\cite{Netz01}, and is supported by numerical simulations \cite{AndrePRL}.
This has also been addressed in simulations on charged rods and charged spheres
in Refs. \cite{AllahyarovPRL,LinsePRL,Lise00,Arnold-Naji}.

Finally, as discussed before, specific excluded-volume interactions between 
particles can be accounted for systematically within the virial expansion, Eq. (\ref{eq:ZSC}),
which allows for treating counterions of finite diameter,
$\sigma_c$, within the strong-coupling scheme. However, as clearly seen from 
Eq. (\ref{eq:FESC}), only macroion-counterion excluded-volume interaction 
enters in the leading-order contribution to the free energy
in the SC limit. The excluded-volume interaction 
between counterions themselves appears only in the higher-order terms, 
which are not considered here. For the sake of simplicity 
in the forthcoming calculations, we shall 
assume a hard-core repulsion between counterions 
and macroions, therefore, overlapping configurations do not 
contribute to the spatial integral in Eq. (\ref{eq:FESC}). In other words,
the effect of counterion size in the SC limit is incorporated 
into the geometry function ${\tilde \Omega}$, which, 
for cylindrical and spherical macroions of radius $R_0$, implies 
a hard-core radius of
\begin{equation}
    R=R_0+\sigma_c/2.
\label{eq:Reff}
\end{equation}
Now, for a given amount of macroion charge, Equation (\ref{eq:Reff})
implies a reduced surface charge density, and thus
an increased Gouy-Chapman length, Eq. (\ref{eq:GC}), as compared to
the case with point-like counterions. 
The following results are, however, presented in units of 
the Gouy-Chapman length
and the counterion diameter explicitly appears only when the actual units
are restored (see {\em e.g.} Eqs. (\ref{eq:Delta_rods}) 
and (\ref{eq:Delta_sph}) below).

\begin{figure}[t]
\begin{center}
\includegraphics[angle=0,width=6.4cm]{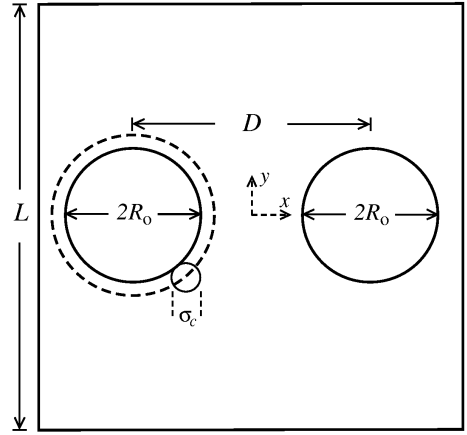}
\smallskip
\caption{\label{fig:tworods} 
The geometry of the system of two similar rods confined in a 
rectangular box (top view) as considered in the text (Section \ref{sec:rods}).
Dashed circle shows the closest approach distance between macroions of 
radius $R_0$ and counterions of diameter $\sigma_c$ (see Eq. (\ref{eq:Reff})).}
\end{center}
\end{figure}




\section{Two Like-Charged Rods}
\label{sec:rods}

Consider two similar and infinitely-long rods each of
radius $R_0$, length $H$ and the linear charge density $\tau$
(in unit of the elementary charge $e$). 
We assume that the axes of the rods are parallel 
and located at separation  $D$ from each other. 
(We choose the frame of coordinates such that its origin lies 
in the mid-way between the rods and its
$z$-axis is parallel to the rods axes -- see Figure \ref{fig:tworods}.)
To consider a more general case, we suppose that
in lateral directions, the system is confined in a rectangular box
with the edge size of $L$.
The electroneutrality condition is fulfilled globally due to the presence 
of oppositely-charged counterions of valency $q$ inside the box.
Assuming that the electric charge is distributed uniformly over the
surface of the rods, their surface charge density reads
$\sigma_s=\tau/(2\pi R)$, and the 
Gouy-Chapman length, Eq. (\ref{eq:GC}), is
\begin{equation}
    \mu=\frac{R}{\ell_B q \tau},
\label{eq:mu_rods}
\end{equation}
where $R$ is the hard-core radius defined according to Eq. (\ref{eq:Reff}).
As it follows from Eq. (\ref{eq:mu_rods}), 
the rescaled rod radius, ${\tilde R}=R/\mu$, is
identical with the {\em Manning parameter}, {\em i.e.}
\begin{equation}
  {\tilde R}=\xi=\ell_B q \tau.
  \label{eq:Rrods}
\end{equation}

The rescaled SC free energy of the two-rod system,
${\mathcal F}_1$,
follows from Eq. (\ref{eq:FESC}), in which the zero-particle
interaction energy, ${\tilde U}_0$,  and the one-particle interaction 
energy,
${\tilde u}$, are obtained from 
Eqs. (\ref{eq:U0}) and (\ref{eq:u}) respectively. 
Neglecting irrelevant self-energy terms and taking the off-set point 
${\mathbf r}_0={\mathbf 0}$, we have
\begin{eqnarray}
 \frac{{\tilde U}_0}{\tilde H}&=&6\pi{\tilde R}^2\ln {\tilde D},
        \\
  {\tilde u}({\tilde x},{\tilde y})&=&
      2{\tilde R}\,[\ln {\tilde r}_1+\ln {\tilde r}_2 
       -\ln ({\tilde D}^2/4)],
\end{eqnarray}
where 
\begin{eqnarray}
     {\tilde r}_1 &=& 
		[({\tilde x}+{\tilde D}/2)^2+{\tilde y}^2]^{1/2},
		\nonumber\\ 
     {\tilde r}_2 &=&
		[({\tilde x}-{\tilde D}/2)^2+{\tilde y}^2]^{1/2},
\label{eq:r1}
\end{eqnarray} 
are radial distances from the rods axes.  
Inserting the above expressions into Eq. (\ref{eq:FESC}), 
we end up with the SC free energy (per unit length of the rods,
${\tilde H}$, and up to an irrelevant additive term) as
\begin{eqnarray}
  \frac{{\mathcal F}_1}{\tilde H} = -2 {\tilde R}^2\ln {\tilde D}  
                           -2{\tilde R}\ln I, 
 \label{eq:F1rods}
\end{eqnarray}
where
\begin{equation}
    I({\tilde D}, {\tilde R}, {\tilde L})\equiv
            \int\,d{\tilde x}d{\tilde y}\,
                          {\tilde \Omega}\,e^{-2{\tilde R}
                            [\ln {\tilde r}_1+\ln {\tilde r}_2]}.
\label{eq:Irods}
\end{equation} 
The geometry function 
${\tilde \Omega}={\tilde \Omega}({\tilde x},{\tilde y};{\tilde D},{\tilde R},{\tilde L})$ 
specifies the region of integration,  
{\em i.e.} it is one inside the volume surrounded  by
the rectangular box and the cylindrical rods, 
and is zero elsewhere.


\subsection{Threshold of attraction}
\label{subsec:thre_rods}

The first term in the rescaled SC free energy, Eq. (\ref{eq:F1rods}), 
simply gives the long-ranged (bare)
electrostatic repulsion between the rods, and the second 
term involving the spatial integral contains energetic and entropic 
contributions of counterions on the {\em leading order}, which can generate
an effective rod-rod attraction. In order to examine the counterionic
contribution, and the onset of attraction, 
let us consider first the limit of two unconfined rods, {\em i.e.}
 when the box size becomes infinitely larger than the rod radius,
${\tilde L}/{\tilde R}\rightarrow \infty$. In this limit, 
the counterionic integral $I$, Eq. (\ref{eq:Irods}), 
diverges for Manning parameters ${\tilde R}<1/2$, which 
can be seen by rescaling the spatial coordinates with the box size
according to ${\tilde x}\rightarrow {\tilde x}/{\tilde L}$, {\em etc.}, 
that yields $I\sim {\tilde L}^{2-4{\tilde R}}$. 
This indicates that ${\tilde R}=1/2$ is the 
threshold of counterion condensation in this system, below which
the distribution function of counterions around the rods vanishes 
\cite{Note3} in agreement with the value obtained from Manning 
condensation theory \cite{Manning97}. Moreover, 
the counterion-mediated force,  
\begin{equation}
{\tilde F}_{ci}\sim \frac{\partial}{\partial {\tilde D}} \ln I, 
\end{equation}
vanishes for ${\tilde R}<1/2$ as the box size tends to infinity. 
Thus, attraction between 
unconfined rods can only set in for Manning parameters larger than 1/2.
Now assuming that the SC free energy has only one local minimum, 
which is indeed the case as we shall see from the numerical solution,
the onset of attraction can be determined by examining the 
large-separation behavior of the free energy, {\em i.e.} for 
${\tilde D}\gg{\tilde D}_{min}$, where ${\tilde D}_{min}=2{\tilde R}$
is the smallest possible axial separation of the rods.
To this end, we rescale the spatial coordinates with the axial separation
${\tilde D}$ as
\begin{equation}
   {\tilde x'}=\frac{\tilde x}{\tilde D}, \,\,\, 
   {\tilde y'}=\frac{\tilde y}{\tilde D}.
\label{eq:xyrescale}
\end{equation}
Accordingly, the
integral $I$ in Eq. (\ref{eq:Irods}) scales as
\begin{equation}
   I({\tilde D},{\tilde R})={\tilde D}^{2-4{\tilde R}}
                          J(\frac{\tilde R}{\tilde D}, {\tilde R}),
\label{eq:IJrods}  
\end{equation}
where $J$ is a dimensionless integral given by 
\begin{equation}
   J=2\int_{{\tilde x'}>0}d{\tilde x'}d{\tilde y'}
                   [({\tilde x'}+\frac{1}{2})^2+({\tilde y'})^{2}]^{-{\tilde R}}
                   [({\tilde x'}-\frac{1}{2})^2+({\tilde y'})^{2}]^{-{\tilde R}}
                   {\tilde \Omega'},            
\label{eq:J}
\end{equation}
where we have made use of the symmetry property of the integrand
upon the reflection with respect to the plane 
${\tilde x'}=0$, and thus the corresponding integral runs over the half-space 
${\tilde x'}>0$ excluding a disk of
radius ${\tilde R}/{\tilde D}$ centered at 
$({\tilde x'}=+1/2, {\tilde y'}=0)$ (this is
formally accounted for by the geometry function ${\tilde \Omega'}$).
For very large ${\tilde D}/{\tilde D}_{min}$, the radius of the disk 
tends to zero and the limiting behavior of the integral in Eq. (\ref{eq:J}) 
is determined by the contributions from the boundary regions, which vary
depending on whether ${\tilde R}$ is smaller or larger than 1.

For Manning parameter
${\tilde R}<1$, the contribution of the boundary region around the disk
vanishes, and the integral in Eq. (\ref{eq:J})
is dominated by its outer boundary, which gives only a constant
independent of ${\tilde D}$. Therefore, the prefactor of $J$ in 
Eq. (\ref{eq:IJrods}) yields the leading ${\tilde D}$-dependence of $I$  
for large axial separations, and substituting this into 
Eq. (\ref{eq:F1rods}) gives the limiting form of 
the rescaled SC free energy as 
\begin{equation}
  \frac{{\mathcal F}_1}{\tilde H} \approx
        -2 {\tilde R}(2-3{\tilde R})\ln {\tilde D}.
\label{eq:asyF1rods}
\end{equation}
 For ${\tilde R}>1$, on the other hand, the integral 
in Eq. (\ref{eq:J})
is dominated by the boundary region around the disk yielding 
$J\sim{\tilde D}^{2{\tilde R}-2}$ for very large 
${\tilde D}/{\tilde D}_{min}$, which leads to the following
 attractive tail for the rescaled SC free energy,
\begin{equation}
  \frac{{\mathcal F}_1}{\tilde H} \approx 2{\tilde R}^2\ln {\tilde D}.
\label{eq:asyF1rods_2}
\end{equation}
Therefore, as clearly seen from Eqs. (\ref{eq:asyF1rods}) 
and (\ref{eq:asyF1rods_2}), two unconfined rods 
experience a repulsive force at large separations,
when Manning parameter, ${\tilde R}$, becomes smaller than the threshold
\begin{equation}
{\tilde R}_c=\frac{2}{3},
\label{eq:R_c}
\end{equation}
and on the contrary, they attract each other for larger Manning parameters
${\tilde R}>{\tilde R}_c$. This result is obtained also by 
numerical calculation of the free energy (Section \ref{subsec:equisep_rods}).

\begin{figure}[t]
\begin{center}
\includegraphics[angle=0,width=7.5cm]{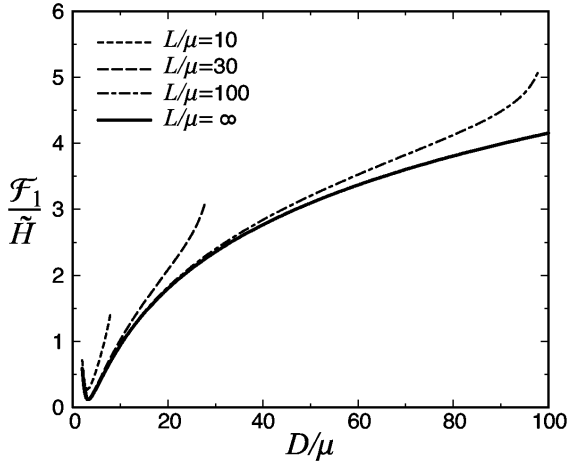}
\smallskip
\caption{\label{fig:freeRlarge}
The rescaled strong-coupling free energy of the two-rod system, 
Eq. (\ref{eq:F1rods}), 
as a function of the rescaled axial separation, ${\tilde D}=D/\mu$,
for Manning parameter ${\tilde R}=1.0$ and different box sizes
indicated on the graph.}
\end{center}
\end{figure}

The result that the onset of attraction for two unconfined rods
is predicted to be larger than the threshold of the 
counterion condensation (${\tilde R}=1/2$) could be anticipated
from the beginning, since right at the threshold of the condensation, 
there remains an unbalanced bare repulsion between the rods according to
the SC free energy, Eq. (\ref{eq:F1rods}). 
The onset of attraction has also been considered
in a number of recent studies. Analysis of Ray and Manning \cite{Manning97} 
based on the standard counterion-condensation model predicts attraction
for ${\tilde R}>1/2$. In contrast, 
results from the counterion-condensation theory of 
Arenzon {\em et al.} \cite{Arenzon99} suggest a larger value of  
${\tilde R}=2$ for the threshold of attraction. Numerical simulations,
on the other hand,  indicate the appearance of 
attraction for a wide range of Manning parameters including ${\tilde R}\approx 1$
\cite{Gron97,Deserno03,Arnold03,Arnold-Naji}. The precise 
location of the threshold of attraction between two charged rods
is currently being studied by 
means of MD simulations \cite{Arnold03,Arnold-Naji}.

The attractive force predicted in the strong-coupling limit
is long-ranged and scales inversely with 
the axial separation between the rods 
(see Eqs. (\ref{eq:asyF1rods}) and (\ref{eq:asyF1rods_2}) and 
Figure \ref{fig:freeRlarge}). 
It should be noted that the SC attraction originates mainly
from the energetic interaction mediated by counterions
sandwiched between the rods, and in this respect, 
is different from the attraction obtained 
by the Gaussian-fluctuation theories
\cite{Oosawa,Barrat,Podgornik98,Ha}.
Specifically, for increasing Manning parameter (or decreasing 
temperature), the strength of SC attraction increases and saturates to a
temperature-independent value (see Appendix \ref{subsec:asymp_rods}).
This can be seen also from the large-separation behavior of the
free energy for highly-charged rods, Eq. (\ref{eq:asyF1rods_2}), 
which gives the attractive force as
\begin{equation}
  \frac{F_{rods}}{H}\approx -\frac{e^2\tau^2}{2\pi\varepsilon\varepsilon_0 D}
\end{equation}
in actual units, where $F_{rods}$ is 
obtained from the actual SC free energy, Eq. (\ref{eq:FSCF1}),
as $F_{rods}=-(k_BT)\partial {\mathcal F}_N^{SC}/\partial D$. 
Such an energetic attraction
is also obtained for planar charged walls \cite{Netz01,AndrePRL} and 
charged spheres (Section \ref{sec:spheres}),
but with different dependencies on $D$. The mechanism of SC attraction 
for large ${\tilde R}$, therefore, qualitatively agrees with the 
low-temperature results \cite{Arenzon99,Shklovs99}
(see also Refs. \cite{Ha,Ha01,Lau01} for discussions on the crossover 
from the low-temperature to the high-temperature regime).

\begin{figure}[t]
\begin{center}
\includegraphics[angle=0,width=7.5cm]{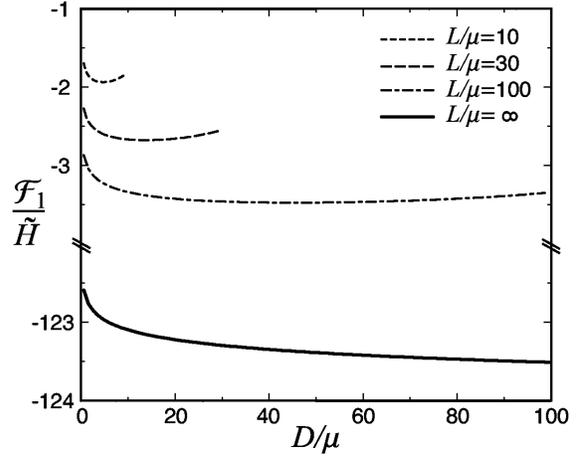}
\smallskip
\caption{\label{fig:freeRsmall}
The same as Figure \ref{fig:freeRlarge}, but for 
Manning parameter ${\tilde R}=0.3$.}
\end{center}
\end{figure}


\subsection{Equilibrium axial distance}
\label{subsec:equisep_rods}

Now, let us consider a more general case with a rectangular box of 
edge size $L$ (Figure \ref{fig:tworods}). For sufficiently large
Manning parameters ${\tilde R}>{\tilde R}_c=2/3$, the presence of 
confining walls is expected to have only a minute effect on the effective
rod-rod interaction since most of the counterions are 
strongly localized in the proximity of the 
rods. Figure \ref{fig:freeRlarge} shows the rescaled  
SC free energy of the two-rod system, 
calculated numerically from Eq. (\ref{eq:F1rods}), 
as a function of the axial separation, ${\tilde D}$, for Manning 
parameter ${\tilde R}=1$ and various box sizes. 
As seen, the free energy quickly converges  to the free energy of an 
unconfined system with ${\tilde L}=\infty$, and 
the long-ranged attraction is only slightly 
strengthened at large separations, when rods are located 
close to the box boundaries. The free energy takes its minimum at 
a small axial distance, ${\tilde D}_\ast\approx 2{\tilde R}$, which 
corresponds to a closely-packed bound state of rods. 
(We may refer to ${\tilde D}_\ast$ as
rescaled ``equilibrium'' axial separation.)
The approximate form of ${\mathcal F}_1$ in the vicinity of its minimum can be 
calculated as 
\begin{equation} 
  \frac{{\mathcal F}_1}{\tilde H} \approx 6 {\tilde R}^2\ln {\tilde D}
              -2 {\tilde R}  \ln ({\tilde D}-2{\tilde R})
 \label{eq:F1largeRrods_txt}
\end{equation} 
for ${\tilde R}\gg 1$ and up to an irrelevant 
additive term (Appendix \ref{subsec:asymp_rods}). 
The first term in Eq. (\ref{eq:F1largeRrods_txt}) 
contributes a dominant (energetic) attractive force and the second
term generates a repulsive force between the rods at small separations.  
Using  Eq. (\ref{eq:F1largeRrods_txt}), the rescaled equilibrium
 axial separation is obtained approximately as
\begin{equation}
   {\tilde D}_\ast\approx 
                 2{\tilde R}+\frac{2}{3}+{\mathcal O}(\frac{1}{{\tilde R}}),
\label{eq:DlargeR_rods_txt}
\end{equation}
for sufficiently large Manning parameter. 
Equation (\ref{eq:DlargeR_rods_txt}) shows that 
the (rescaled) equilibrium surface-to-surface distance of the rods,
${\tilde D}_\ast-2{\tilde R}$, decreases and
tends to a value of the order of unity as Manning parameter tends to 
infinity. Restoring the actual units in Eq. (\ref{eq:DlargeR_rods_txt}),
we obtain $D_\ast\approx 2R+2\mu/3+{\mathcal O}(\mu^2/R)$, where
$R$ and $\mu$ are given by Eqs. (\ref{eq:Reff}) and (\ref{eq:mu_rods}). 
Thus, the actual equilibrium surface-to-surface separation 
is obtained approximately as 
\begin{equation}
  \Delta_\ast\equiv D_\ast-2R_0\approx \sigma_c+\frac{2}{3}\mu+{\mathcal O}(\mu^2),
\label{eq:Delta_rods}
\end{equation}
when the Manning parameter is sufficiently large 
(or the Gouy-Chapman length, $\mu$, is small), that is about the
counterion diameter.

\begin{figure}[t]
\begin{center}
\includegraphics[angle=0,width=7.5cm]{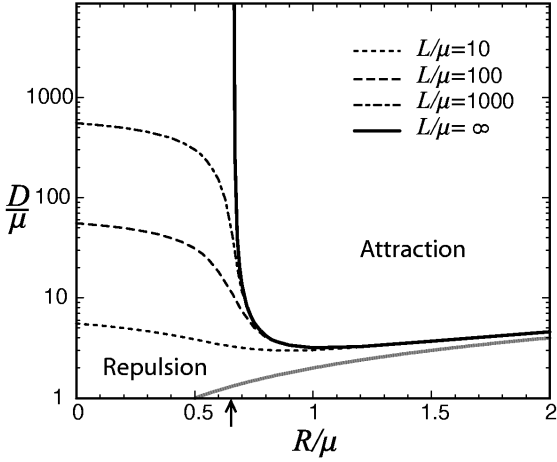}
\smallskip
\caption{
\label{fig:DvsR} 
Rescaled equilibrium axial separation of two like-charged rods
plotted as a function of 
Manning parameter (the rescaled rod radius), ${\tilde R}=R/\mu$, 
for the box sizes 
${\tilde L}=L/\mu=10, 100, 1000, \infty$ as indicated on the graph.  
The small arrow locates the threshold Manning parameter ${\tilde R}_c$=2/3, 
and the thick grey curve corresponds to the contact
axial separation ${\tilde D}_{min}=2{\tilde R}$.}
\end{center}
\end{figure}

For small Manning parameters ${\tilde R}<{\tilde R}_c=2/3$, qualitatively 
different features arise. The effective interaction between the rods 
exhibits a significant
dependence upon the size of the confining box and the 
location of the minimum of the free energy tends to infinity
for ${\tilde L}\rightarrow \infty$ resulting in a pure repulsion
between unconfined rods (see Figure \ref{fig:freeRsmall}). 
This occurs as a result of the dilution
of the counterion cloud around weakly-charged rods. 
For ${\tilde R}\ll 1$, the equilibrium axial separation 
scales with the box size and is obtained approximately 
as (Appendix \ref{subsec:asymp_rods})
\begin{equation}
  {\tilde D}_\ast\approx \frac{{\tilde L}}{\sqrt{\pi}}.
\label{eq:D0_txt}
\end{equation}

\begin{figure}[t]
\begin{center}
\includegraphics[angle=0,width=8.cm]{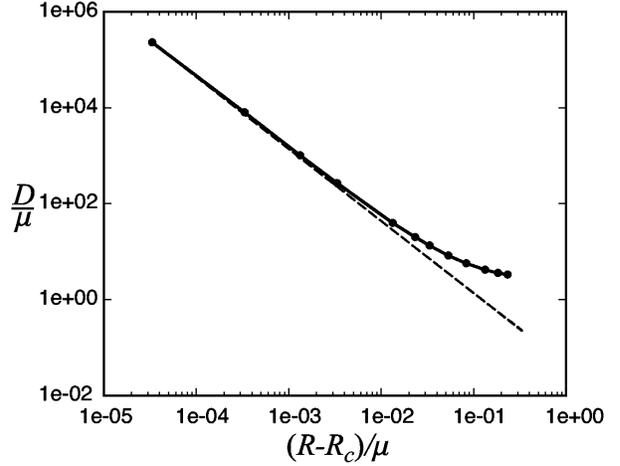}
\smallskip
\caption{
\label{fig:scaling} 
Rescaled 
equilibrium axial separation of
two {\em unconfined} rods (${\tilde L}=\infty$) in the 
vicinity of the threshold Manning parameter $R_c=2/3$ (solid curve
with symbols). The dashed line shows a power-law 
fit with the slope $\alpha=3/2$
(see Eq. (\ref{eq:scaling_rel_rods})).}
\end{center}
\end{figure}

Now, for a given box size, the free energy varies smoothly
from the typical forms shown in Figure 
\ref{fig:freeRlarge} to those shown in Figure \ref{fig:freeRsmall}
by decreasing Manning parameter.
There is, however, a rapid change near ${\tilde R}_c=2/3$
indicative of  a continuous unbinding transition
in an infinitely large box.
This is seen more clearly in Figure \ref{fig:DvsR}, where 
the predictions of the SC theory for the equilibrium axial 
separation is shown as a function of Manning parameter for various 
box sizes. (These results have been obtained directly by numerical
minimization of the free energy, Eq. (\ref{eq:F1rods}).)
The region above a given curve 
corresponds to the axial separations
at which the rods attract, and the region below that corresponds
to the repulsion regime.
When the box size tends to infinity,   
the equilibrium axial separation diverges 
for ${\tilde R}\rightarrow{\tilde R}_c^{+}$ revealing the power-law behavior 
\begin{equation}
    {\tilde D}_\ast\sim ({\tilde R}-{\tilde R}_c)^{-\alpha}
\label{eq:scaling_rel_rods}
\end{equation} 
shown in Figure \ref{fig:scaling}, where within our numerical errors 
\begin{equation}
   \alpha = 3/2.
\label{eq:scaling}
\end{equation}

\subsection{The regime of applicability of SC results}
\label{subsec:sc_valid_rods}

The system of two like-charged rods has been investigated 
in recent numerical simulations revealing strong attractive forces 
 at moderate to large Manning parameters 
\cite{Gron97,Deserno03,Arnold03,Arnold-Naji}. The SC theory provides
quantitative predictions for the equilibrium axial separation, 
that may be compared with simulations results. However, such comparison
with realistic systems of finite coupling parameter 
({\em e.g.} $\Xi\sim 10-100$ for simulations in
Refs. \cite{Gron97,Deserno03,Arnold03,Arnold-Naji}) 
must be performed in a certain regime of parameters, 
where the present asymptotic theory (strictly valid only for
$\Xi\rightarrow \infty$) is applicable. 
To this aim, extensive simulations have been
performed recently by Arnold and Holm \cite{Arnold03} covering a wide range of
parameters, which allow for examining the influence of higher-order 
electrostatic and excluded-volume effects.
The comparison with these results will be presented in separate 
publications \cite{Arnold03,Arnold-Naji}, and in the following,
we merely introduce two criteria identifying the regime of parameters, where
a reasonable agreement between simulations and SC predictions is obtained.

For highly-charged rods, excluded-volume interactions between counterions
may be significant as attraction is accompanied 
by accumulation of counterions in a narrow region between the rods 
\cite{Arnold03} (Appendix \ref{subsec:asymp_rods}). In this situation, the
typical distance between intervening counterions (lining up in the 
$z$-direction), $a_z$,  may be estimated from the local 
electroneutrality condition, $q=2\tau a_z$, giving
\begin{equation}
  a_z=\frac{q}{2\tau},
\label{eq:az_rod}
\end{equation}
where $q$ is the counterion valency and $\tau$ is the single-rod
linear charge density. When, the counterion 
diameter, $\sigma_c$, is larger than $a_z$,  
\begin{equation}
  a_z<\sigma_c,
\end{equation} 
the excluded-volume repulsions between counterions become important, and
further accumulation of counterions between the rods 
according to the SC mechanism is 
prohibited. As a result, the equilibrium separation of the rods 
(determined from the {\em electrostatic component} of the 
total force between them) 
appears to be larger than the present predictions, since the strength
of the mediated attraction drops. Though, it should be
noted that volume interactions may indeed contribute an additional 
attractive component to the total force 
for highly-charged systems \cite{Deserno03,Arnold03}.
It is a priori not clear which of these effects is stronger.

On the other hand, as frequently quoted in the literature 
\cite{AllahyarovPRL,LinsePRL,Lise00,Rouzina96,Netz01,AndrePRL}, 
electrostatic correlations induced by counterions between 
two apposing macroions are
dominant, when the typical distance
 between counterions, $a_z$,
 becomes larger than the macroions surface-to-surface  
distance, $\Delta=D-2R_0$, {\em i.e.} when 
\begin{equation}
\Delta<a_z, 
\label{eq:crit_rods}
\end{equation}
where $a_z$ is given by Eq. (\ref{eq:az_rod}).  
In rescaled units, this condition reads 
\begin{equation}
{\tilde \Delta}<\frac{\Xi}{2\xi},
\end{equation} 
where the coupling parameter, Eq. (\ref{eq:Xi}), is
\begin{equation}
   \Xi = \frac{q^3\ell_B^2\tau}{R}
\end{equation}
for charged rods. The above criterion identifies 
the regime of parameters, where
basically the counterion-rod correlations
are superior to the counterion-counterion correlations
and the present asymptotic theory is applicable
\cite{Arnold-Naji,Netz01,AndrePRL}. 
For moderate coupling parameters $\Xi\sim 10$, this regime can be reached
using charged rods with Manning parameter larger than
$ \xi \approx 1$ \cite{Arnold-Naji}.




\section{Two Like-Charged Spheres}
\label{sec:spheres}

Consider a system of two like-charged 
spheres each of radius $R_0$ and charge valency $Z$. 
The spheres are located at a center-to-center separation of
$D$, and together with their neutralizing counterions are 
confined in a {\em cubic} box of edge size $L$. 
(The geometry of the system is similar to
what we have sketched in Figure \ref{fig:tworods}, where
the two solid circles now display
the largest cross-section of the spheres.) 
We assume the electric charge to be distributed 
uniformly over the surface of the spheres, thus their  
surface charge density reads $\sigma_s=Z/(4\pi R^2)$, and the Gouy-Chapman 
length, Eq. (\ref{eq:GC}), is
\begin{equation}
    \mu=\frac{2R^2}{\ell_B q Z},
\label{eq:mu_sph}
\end{equation} 
where $R$ is the (hard-core) radius of spheres, Eq. (\ref{eq:Reff}). 
In rescaled units, we have 
\begin{equation}
 {\tilde R}=\frac{R}{\mu}=\frac{\ell_B q Z}{2R},
 \label{eq:Rsph}
\end{equation}
which is referred to in the following as {\em Manning parameter} 
for charged spheres in analogy with the cylindrical case, 
Eq. (\ref{eq:Rrods}).

\begin{figure*}[t]
\begin{center}
\includegraphics[angle=0,width=14cm]{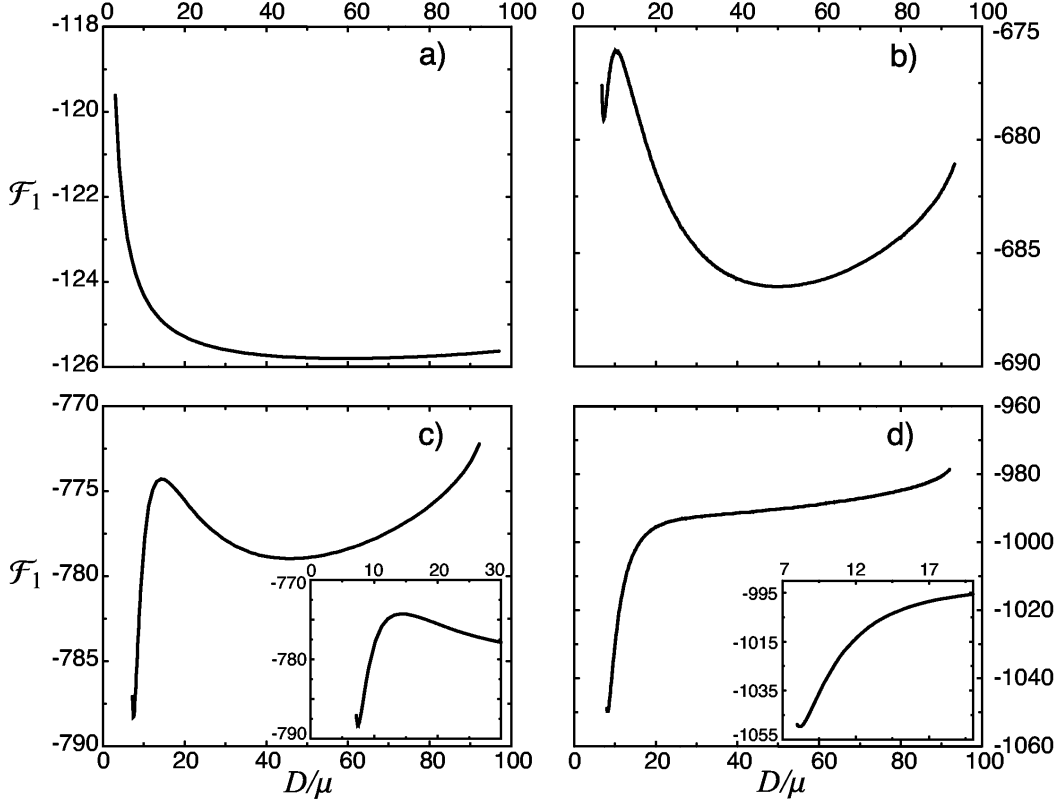}
\smallskip
\caption{
\label{fig:freeS} 
The rescaled SC free energy of the two-sphere system, Eq. (\ref{eq:F1sph}),
plotted as a function of the rescaled center-to-center distance
for different Manning parameters a) ${\tilde R}=1.5$, b) ${\tilde R}=3.4$,
c) ${\tilde R}=3.6$ and d) ${\tilde R}=4.0$. The size of the confining
box is ${\tilde L}=100$ in rescaled units. The insets show a closer view of
the local minimum at ${\tilde D}\approx 2{\tilde R}$.}
\end{center}
\end{figure*}

To study the strong-coupling interaction of the two spheres,
we follow similar lines
as presented 
in the preceding Section by calculating the rescaled SC free energy, 
Eq. (\ref{eq:FESC}).
The zero-particle and one-particle interaction energies
are obtained from Eqs. (\ref{eq:U0}) and (\ref{eq:u}) respectively as 
\begin{eqnarray}
   {\tilde U}_0&=&-28\pi\frac{{\tilde R}^4}{{\tilde D}},
     \nonumber\\
       {\tilde u}({\tilde x},{\tilde y},{\tilde z})&=&
      -2{\tilde R}^2\,[\frac{1}{{\tilde r}_1}+\frac{1}{{\tilde r}_2}
                      -\frac{4}{\tilde D}],
\label{eq:U0u_sph}
\end{eqnarray}
where  
\begin{eqnarray}
      {\tilde r}_1 &=& 
          [({\tilde x}+{\tilde D}/2)^2+{\tilde y}^2+{\tilde z}^2]^{1/2},
      \nonumber\\
      {\tilde r}_2 &=& 
         [({\tilde x}-{\tilde D}/2)^2+{\tilde y}^2+{\tilde z}^2]^{1/2},
\end{eqnarray} 
are radial distances from the centers of the spheres. The reference point
is arbitrarily chosen to be in the mid-way between the spheres 
${\mathbf r}_0={\mathbf 0}$, and self-energy terms are neglected. 
Using Eqs. (\ref{eq:FESC}) and (\ref{eq:U0u_sph}), the rescaled SC free 
energy is obtained as
\begin{eqnarray}
  {\mathcal F}_1 = 4\frac{{\tilde R}^4}{{\tilde D}}
                           -4{\tilde R}^2
                                \ln I,
\label{eq:F1sph}
\end{eqnarray}
where 
\begin{equation}
  I({\tilde D},{\tilde R},{\tilde L})\equiv
                   \int\,d{\tilde x}d{\tilde y}d{\tilde z}\,
                               {\tilde \Omega}\,
                              e^{2{\tilde R}^2\,
                               [\frac{1}{{\tilde r}_1}+
                                 \frac{1}{{\tilde r}_2}]}.
\label{eq:I_sph} 
\end{equation}
The geometry function 
${\tilde \Omega}({\tilde x},{\tilde y},{\tilde z};{\tilde D},{\tilde R},{\tilde L})$ 
specifies the region accessible for counterions, {\em i.e.} it is one inside
the volume surrounded by the cubic box and the two spheres, and is zero elsewhere.

The first term in  Eq. (\ref{eq:F1sph}) gives the long-ranged bare
repulsion between the spheres, and the second 
term involving the spatial integral, $I$,
contains energetic and entropic 
contributions of counterions on the leading order
reproducing the counterion-condensation process and an attractive force 
for highly-charged spheres.

\begin{figure}[t]
\begin{center}
\includegraphics[angle=0,width=7.5cm]{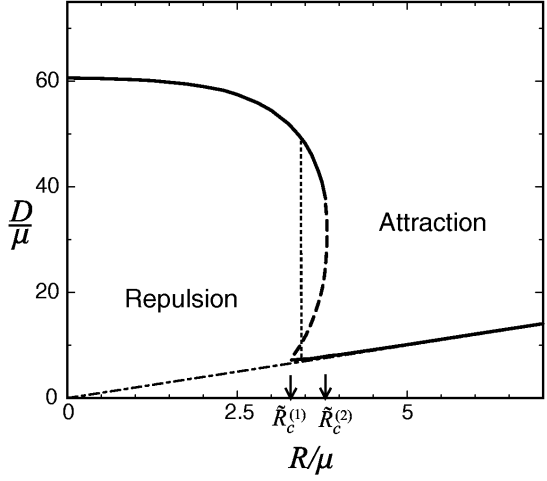}
\smallskip
\caption{
\label{fig:DminimalS} 
Locations of the minima (solid curves) and the maximum (dashed curve)
of the SC free energy for the two-sphere system plotted 
as a function of Manning parameter ${\tilde R}$
for ${\tilde L}=100$. 
The two arrows  on the horizontal axis 
show the locations of the threshold Manning parameters
${\tilde R}_c^{(1)}$ and ${\tilde R}_c^{(2)}$. 
The dot-dashed line represents 
the contact separation ${\tilde D}_{min}=2{\tilde R}$, and
the vertical dotted line shows the Manning parameter for which
the values of the two minima of the free energy are equal.   
}
\end{center}
\end{figure}

It follows easily from Eqs. (\ref{eq:F1sph}) and (\ref{eq:I_sph}) that for finite Manning 
parameter, the counterion-mediated force vanishes as the size of 
the confining box tends to infinity resulting in pure repulsion between 
unconfined spheres. This  goes back to the 
fact that in the absence of confining boundaries, spherical macroions can 
not bind counterions that tend to diffuse in order to gain entropy.  
This is reflected by the divergency of the counterionic integral, 
Eq. (\ref{eq:I_sph}), as ${\tilde L}\rightarrow \infty$. 
(Note that the involved integrand is always 
positive and larger than one, and upon rescaling the spatial coordinates 
as ${\tilde{\mathbf r}}'={\tilde{\mathbf r}}/{\tilde L}$, it
scales with the volume of the confining box as $I\sim {\tilde L}^3$.) 
As a result, the distribution function of
counterions \cite{Note3} as well as the counterion-mediated force,
\begin{equation}
  {\tilde F}_{ci}\sim \frac{\partial}{\partial {\tilde D}} \ln I,
\end{equation}
that scales with the box size as $\sim {\tilde L}^{-2}$, vanish
for increasing box size. In contrast,
it turns out that two like-charged spheres can
attract each other in {\em any} finite box
provided that their Manning parameter is
sufficiently large as will be demonstrated below.

Figures \ref{fig:freeS}a-d show the typical form of the SC free energy, 
calculated numerically (using Monte-Carlo integration methods) from 
Eq. (\ref{eq:F1sph}), for two spheres in  a box with 
${\tilde L}=100$ and for several 
Manning parameters. As seen, at sufficiently 
small Manning parameter, the long-ranged repulsion is dominant 
and only a weak attraction operates, when the spheres are at large
separations comparable with the box size (Figure \ref{fig:freeS}a). But at
Manning parameters larger than a threshold ${\tilde R}_c^{(1)}$ 
(which is obtained as ${\tilde R}_c^{(1)}\approx 3.3$ for 
${\tilde L}=100$), a local minimum (short-ranged attraction) 
emerges at small separations about
${\tilde D}\approx 2{\tilde R}$ corresponding to a
meta-stable bound state of spheres (Figure \ref{fig:freeS}b). 
The attractive interaction develops further and the potential barrier
disappears, when the Manning parameter becomes 
larger than the second threshold ${\tilde R}_c^{(2)}$ (which is
${\tilde R}_c^{(2)}\approx 3.8$ for ${\tilde L}=100$)
-- see Figure \ref{fig:freeS}d. This
clearly demonstrates the occurrence of a {\em discontinuous} 
transition between a closely-packed bound state and a repulsion-dominated
state of two spheres by changing Manning parameter ${\tilde R}$.

In Figure \ref{fig:DminimalS}, we show the locations of the minima (solid curves) 
and the maximum (dashed curve) of the SC free energy as a function of 
Manning parameter for ${\tilde L}=100$ (obtained numerically using 
Eq. (\ref{eq:F1sph})). The upper branch represents the 
shallow repulsion-dominated minimum, which is strongly
sensitive to the box size. For small Manning parameter ${\tilde R}\ll 1$, 
the corresponding equilibrium center-to-center distance increases linearly 
with the box size and is obtained approximately as 
\begin{equation}
  {\tilde D}_\ast\approx \sqrt[3]{\frac{3}{4\pi}} {\tilde L},
\label{eq:D0sph}
\end{equation} 
for ${\tilde L}\rightarrow \infty$ (Appendix \ref{subsec:asymp_sph}).
By contrast, the small-separation minimum (the lower branch)
is effectively independent of the box size and is maintained essentially 
by an attractive interaction mediated by  counterions 
intervening between the spheres. For sufficiently large Manning parameter
${\tilde R}\gg 1$, the approximate form of the rescaled SC 
free energy around its minimum at small separations is obtained as 
\begin{equation}
  {\mathcal F}_1 = -28\frac{{\tilde R}^4}{\tilde D}
       -4{\tilde R}^2  \ln ({\tilde D}-2{\tilde R})
\label{eq:F1sphRlarge_txt}
\end{equation} 
up to an unimportant additive term (Appendix \ref{subsec:asymp_sph}). 
For increasing Manning parameter 
(or decreasing temperature), the free energy expression (\ref{eq:F1sphRlarge_txt}) 
leads to an (energetic) attractive force independent of temperature, which 
scales with the center-to-center distance of the 
spheres, $D$, as $\sim 1/D^2$, in agreement with  results obtained by Shklovskii 
\cite{Shklovs99}. Using  Eq. (\ref{eq:F1sphRlarge_txt}), the rescaled equilibrium
center-to-center distance is obtained approximately as
\begin{equation}
   {\tilde D}_\ast\approx 2{\tilde R}+\frac{4}{7}+{\mathcal O}(\frac{1}{{\tilde R}}),
\label{eq:DlargeRsph}
\end{equation}
which yields an equilibrium surface-to-surface distance,
${\tilde D}_\ast-2{\tilde R}$, of the order of unity in rescaled units.
Now restoring the actual units in Eq. (\ref{eq:DlargeRsph}),
we obtain $D_\ast\approx 2R+4\mu/7+{\mathcal O}(\mu^2/R)$,
where $R$ and $\mu$ are given by Eqs. (\ref{eq:Reff}) and (\ref{eq:mu_sph}). 
Thus, the actual equilibrium surface-to-surface separation of spheres 
is obtained approximately as
\begin{equation}
  \Delta_\ast\equiv D_\ast-2R_0\approx \sigma_c+\frac{4}{7}\mu+{\mathcal O}(\mu^2),
\label{eq:Delta_sph}
\end{equation}
when the Manning parameter is sufficiently large 
(or the Gouy-Chapman length, $\mu$, is small), that is about the
counterion diameter.

\begin{figure}[t]
\begin{center}
\includegraphics[angle=0,width=7.5cm]{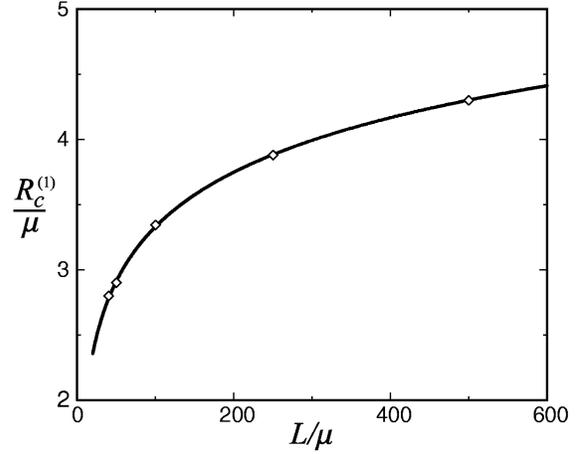}
\smallskip
\caption{
\label{fig:Rc1} 
The threshold Manning parameter ${\tilde R}_c^{(1)}$ 
plotted as a function of the box size ${\tilde L}$ 
(see Eq. (\ref{eq:Rc1}) in the text). This threshold 
is obtained for box sizes larger than ${\tilde L}\approx 20$.}
\end{center}
\end{figure}

\begin{table*}[t]
\begin{center}
\begin{tabular*}{18.cm}{lcccccc|ccccc|cc} 
\hline\hline
Simulation & $q$ & $Z$ & ${\ell_B}$(\AA) & $R_0$(\AA) & $\sigma_c$(\AA) & 
$L$(\AA) & $\mu$(\AA) & ${\tilde R}$  & $\Xi$ & ${\tilde R}_c^{(1)}$  
& $a_\bot$(\AA) & $\Delta_{sim}$(\AA) & $\Delta_\ast$(\AA)
\\ \hline
Gr{\o}nbech-Jensen   & 2 & 10 & 7.01 & 7& 3.3& 
50-200& 1.07 & 8.1 & 26  & 2.8-3.7  & 7.7 & 2.5 & 3.41   \\
{\em et al.} \cite{Gron98} &&&&&&&&&&&&& \\
Wu {\em et al.} \cite{Wu} & 2 & 20 & 7.14 & 10 & 4 & 100 & 1.01 & 11.9  & 28  &  
3.3 & 7.5 &  4 & 4.10 \\  
Allahyarov &2 & 32 & 112 & 48.9 & 4.4 & 
$\sim 10^2$ & 0.73 & 70.1 & 615   & $\sim 3.3$ &  25.5 & -- & 4.41  \\
{\em et al.} \cite{AllahyarovPRL} &&&&&&&&&&&&& \\
Linse {\em et al.} \cite{LinsePRL,Lise00} & 3 & 60 & 7.15 & 20 & 4 & $\sim 
10^2$ & 0.75 & 29.2 & 85  & $\sim 3.3$ & 9.8 &  4 & 4.05  \\
Hribar  {\em et al.} \cite{Hribar} & 3 & 12 & 7.15 & 10 & 2 & $\sim 10^2$ & 
0.94 & 11.7 & 68 & $\sim 3.3$  & 11.0 &  -- & 2.09 \\ 
\hline\hline 
\end{tabular*}
\end{center}
\caption{\label{tab:sim_parameters_sph} Parameters from simulations on 
highly-charged spheres (see the text for the
definitions). The last two columns show the equilibrium 
surface-to-surface distance, $\Delta_{sim}$, obtained
in these simulations (if explicitly measured), and
the strong-coupling prediction, $\Delta_\ast=D_\ast-2R_0$. 
Some of the numbers are given up to the order of magnitude, and the 
extracted values of $\Delta$ from simulations
have a typical resolution of about 1\AA.}
\vspace*{5mm}
\end{table*}

Similar features are obtained for 
box sizes that are
larger than roughly ${\tilde L}\approx 20$; for 
smaller box size only the shallow large-distance minimum is obtained as depicted 
in Figure \ref{fig:freeS}a. 
The striking result is that the threshold Manning parameters
${\tilde R}_c^{(1)}$ and ${\tilde R}_c^{(2)}$
monotonically increase by increasing the confinement volume. 
This behavior is shown in Figure \ref{fig:Rc1} for 
${\tilde R}_c^{(1)}$ (symbols), along with  the best  
logarithmic fit to these results (solid curve) as
\begin{equation}
  {\tilde R}_c^{(1)}=a+b\ln {\tilde L},
\label{eq:Rc1}
\end{equation}
where $a\approx 0.55$ and $b\approx 0.6$. Such a weak
increase of the attraction 
threshold demonstrates that even in a very large confinement, 
two like-charged spheres can fall into the attraction regime provided
that the corresponding Manning parameter exceeds a {\em moderate} threshold. 
This behavior may also explain the stability of closely-packed
clusters  of highly-charged spheres 
in large confinement, and their insensitivity to the box size as 
addressed in recent simulations \cite{Gron98,AllahyarovPRL,LinsePRL,Lise00}. 
The simulation parameters, where strong attraction has been observed, 
indeed covers the predicted attraction-dominated regime 
${\tilde R}\gg {\tilde R}_c^{(1)}$ (see Table \ref{tab:sim_parameters_sph}).

\begin{figure}[t]
\begin{center}
\includegraphics[angle=0,width=7.5cm]{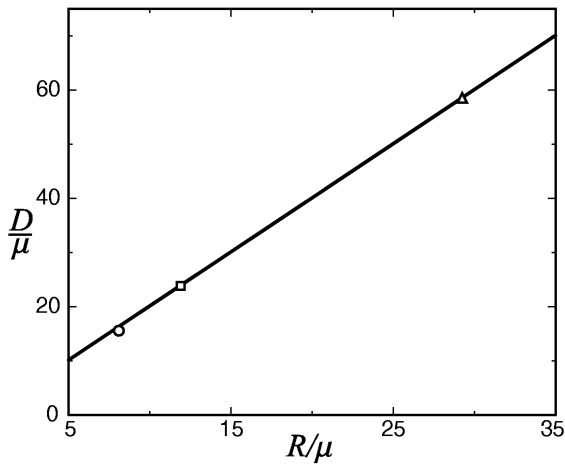}
\smallskip
\caption{
\label{fig:simSph} 
Rescaled equilibrium center-to-center separation of like-charged spheres
as a function of Manning parameter in the attraction-dominated regime 
({\em i.e.} ${\tilde R}\gg {\tilde R}_c^{(1)}$).
Symbols show data obtained from recent 
simulations (circle: Gr{\o}nbech-Jensen {\em et al.} \cite{Gron98},
square: Wu {\em et al.} \cite{Wu}, triangle: Linse {\em et al.} 
\cite{LinsePRL,Lise00}) compared with the SC prediction
(solid line).}
\end{center}
\end{figure}

\begin{figure}[t]
\begin{center}
\includegraphics[angle=0,width=7.5cm]{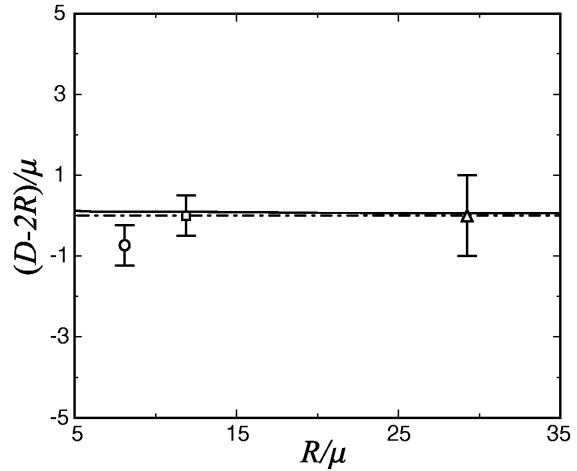}
\smallskip
\caption{
\label{fig:simSph_diff} 
Rescaled equilibrium surface-to-surface separation of like-charged spheres
as a function of Manning parameter in the attraction-dominated regime.
Symbols show data from recent 
simulations (circle: Gr{\o}nbech-Jensen {\em et al.} \cite{Gron98},
square: Wu {\em et al.} \cite{Wu}, triangle: Linse {\em et al.} 
\cite{LinsePRL,Lise00}) and the solid line is the SC prediction. 
Dot-dashed line shows the contact separation 
${\tilde D}_{min}=2{\tilde R}$.}
\end{center}
\end{figure}

One should note that the unbinding transition can occur also 
by changing the confinement volume for a given Manning parameter.
This is implied by the fact that, for a given Manning parameter between 
the two thresholds ${\tilde R}^{(1)}_c<{\tilde R}<{\tilde R}^{(2)}_c$, 
the height of the free energy barrier
decreases slowly by increasing the box size. Qualitatively similar features have
been obtained in the simulations by Gr{\o}nbech-Jensen {\em et al.} 
\cite{Gron98} on a system of two like-charged spheres with divalent 
counterions. The obtained potential of mean force between spheres 
was shown to have a local minimum at small 
separation (the compact state) separated by a potential barrier
from a repulsion regime (see Fig. 1 in Ref. \cite{Gron98}, 
which schematically resembles Figure \ref{fig:freeS}. 
Note that the tail of the effective interaction
is influenced by the boundary conditions, that is 
periodic in the case of the simulations).  
The barrier height was found to decrease for increasing confinement 
volume resulting in repulsion between spheres (Fig. 3 in Ref. \cite{Gron98}).
The appearance of a potential barrier between two highly-coupled spheres
was also noted in Ref. \cite{Messina00}.

The strong-coupling regime has also been
investigated in several other simulations 
both on many-sphere systems \cite{LinsePRL,Lise00,Hribar} 
and on a single pair of spheres \cite{Wu,AllahyarovPRL}.
Table \ref{tab:sim_parameters_sph} presents typical parameters from some of
recent simulations. In the two last columns of the Table, we compare the 
equilibrium surface-to-surface separation between attracting
spheres from these simulations, $\Delta_{sim}$ 
(if explicitly estimated), with  the SC prediction, 
$\Delta_\ast=D_\ast-2R_0$, in actual units. 
(In Figures \ref{fig:simSph} and \ref{fig:simSph_diff}, we also show 
the center-to-center distance, ${\tilde D}_\ast$, 
and the surface-to-surface distance, 
${\tilde D}_\ast-2{\tilde R}=(D_\ast-2R_0-\sigma_c)/\mu$, in rescaled units.)
The strong-coupling results have been calculated
numerically from the free energy, Eq. (\ref{eq:F1sph}), and
involve the finite size of counterions, 
which as discussed in Section \ref{subsec:SCfree}, enters at 
the leading order via a finite closest approach distance between 
spheres and counterions. As seen, there is a qualitative agreement between 
the theoretical predictions and the simulations results. 
In fact, as addressed in previous numerical studies (see for example the Discussion 
in Ref. \cite{Lise00}), the equilibrium  surface-to-surface separation of 
spheres in these simulations appears to be of the order of
the counterion diameter.
This result agrees with the SC prediction for highly-charged spheres 
of small Gouy-Chapman length ($\mu\sim 1$\AA)
as demonstrated by the limiting expression (\ref{eq:Delta_sph}).

The regime of applicability of SC predictions to systems with finite coupling strength,
such as the above simulations, may be 
examined by considering the ratio of the lateral counterion 
separation at charged spheres, $a_\bot$, to the surface-to-surface 
distance between the spheres, $\Delta=D-2R_0$
\cite{AllahyarovPRL,LinsePRL,Lise00,Rouzina96,Netz01,AndrePRL}.  
The typical value of $a_\bot$ may be estimated 
from the local electroneutrality condition $\pi a_\bot^2\sim q/\sigma_s$, 
which holds up to a geometric factor of the order of one. In rescaled units, 
we get $a_\bot/\mu= {\tilde a}_\bot \sim (2\Xi)^{1/2}$, where the coupling
parameter, Eq. (\ref{eq:Xi}), reads
\begin{equation}
  \Xi = \frac{q^3\ell_B^2Z}{2R^2}
\label{eq:Xisph}
\end{equation}
for charged spheres.
The lateral separation between counterions may also be written as 
\begin{equation}
  a_\bot\sim R\sqrt{\frac{4q}{Z}}.
\label{eq:az_sph}
\end{equation}
As discussed before, the SC regime, 
{\em i.e.} when higher-order corrections
are small (Section \ref{sec:SCgeneral}), is characterized by  
the condition 
\begin{equation}
  \Delta < a_\bot.
\label{eq:crit_sph}
\end{equation} 
The estimated values of $a_\bot$ for the given  
simulations are shown in Table \ref{tab:sim_parameters_sph}. 
As seen, the above criterion is fulfilled for the simulation 
parameters \cite{AllahyarovPRL,LinsePRL,Lise00}.
In addition, the size of counterions are typically 
smaller than  their lateral separation
indicating that excluded-volume repulsion between counterions, 
which enters through higher-order corrections to the SC theory, 
is not a dominant effect.




\section{Conclusion and Discussion}
\label{sec:discussion}

In this paper, we have presented the predictions of the strong-coupling 
theory for the effective interaction between two like-charged rods
and two like-charged spheres.
The SC theory obtained from a systematic virial expansion 
(Section \ref{sec:SCgeneral}) provides a framework to study 
strongly-coupled systems ($\Xi\gg 1$). 
For large Manning parameters (or weakly-curved macroion surfaces), 
counterions  effectively condense around macroions and 
a long-ranged attraction is obtained. 
(For charged rods, attraction is obtained even in an infinitely large
confining box due to a universal counterion-condensation effect in this limit, 
while spheres show attraction only when they are confined in an outer box.) 
In general, the SC attraction is accompanied by strong
accumulation of counterions in the intervening region
between macroions that gives rise to attractive forces, which 
scale with the macroions separation, $D$, as $\sim 1/D$ for two rods, and 
as $\sim 1/D^2$ for two spheres. 
The equilibrium surface-to-surface separation of highly-charged macroions
is then predicted to be of the order of the counterion diameter 
plus a term of the order of the Gouy-Chapman length --see Eqs. (\ref{eq:Delta_rods})
and (\ref{eq:Delta_sph})-- which is in qualitative 
agreement with results obtained from recent simulations
on both charged rods \cite{Arnold03} 
and charged spheres (Section \ref{sec:spheres}). For point-like
counterions, a quantitative agreement has been established with recent 
Molecular Dynamics simulations \cite{Arnold-Naji}.
Some of the features such as the finite-size dependence
of the effective interaction and the appearance of a potential barrier in 
a system of charged spheres are also consistent with the simulation results
\cite{Gron98,Messina00}.

Note that in the simulations, attractive interactions
are found to have a short-ranged character, whereas the SC theory
predicts long-ranged attractions for large Manning parameters.
This difference should be attributed to finite-coupling effects as
in the simulations considered here 
\cite{Gron98,Wu,AllahyarovPRL,LinsePRL,Lise00,Hribar,Deserno03,Arnold03,Arnold-Naji}, 
the coupling parameter is only moderately large, $\Xi\sim 10-10^2$. 
As it was demonstrated for planar charged walls \cite{Netz01,AndrePRL},
for such coupling parameters, 
the strength of electrostatic correlations is substantially decreased when 
charged walls are far from each other and consequently, the range of attraction
is decreased. (In that case, it was shown that the large-distance behavior of 
the effective interaction is described by the PB theory and is purely repulsive).  
Yet, the SC results were shown to remain valid at sufficiently 
small separations, as obtained from the 
criterion explained in the text \cite{Arnold-Naji,Netz01,AndrePRL} 
(see Eqs. (\ref{eq:crit_rods}) and (\ref{eq:crit_sph})).
As we already discussed, this criterion can be satisfied at moderate
to large couplings using moderate values of Manning parameter, 
{\em e.g.} ${\tilde R}>1$ for two charged rods \cite{Arnold-Naji}. 
Experimentally, this regime can be reached using 
multivalent counterions and highly-charged
macroions. For instance, in aqueous solutions of DNA 
(with radius $R_0\approx 10${\AA} and linear charge density  
$\tau\, e\approx 6\,e/nm$), Manning parameter and the coupling parameter are 
as large as ${\tilde R}\approx 8$ and $\Xi\approx 25$ in the presence of
divalent counterions, and as 
${\tilde R}\approx 12$ and $\Xi\approx 80$ for trivalent counterions such as 
spermidine. In colloidal dispersions, an aqueous solution of 
highly-charged surfactant micelles of, for example,   
typical radius $R_0\approx 20${\AA} and charge valency  $Z\approx 60$ represents
Manning parameter and the coupling parameter of the order of
${\tilde R}\approx 30$ and $\Xi\approx 100$ for trivalent counterions. 
It is also noted that recent experimental observation of 
attraction between like-charged colloidal particles 
in Refs. \cite{Grier94,Kepler,Ise}
concerns the regime of small coupling parameters (typically with 
$\Xi\sim 10^{-2}-10^{-1}$ due to the large size of spheres
$\sim 1\, \mu m$). These observations, therefore, should not be 
compared with the strong-coupling results.

We have also examined the predictions of the SC theory for 
small Manning parameter (or small radius of curvature), 
{\em e.g.} for the onset of attraction. For sufficiently 
small Manning parameters and increasing confinement volume, 
one expects that counterions completely de-condense leading to 
pure repulsion between macroions. It should be noted, however, 
that when counterions de-condense, counterion-macroion correlations 
become effectively small and contributions from higher-order 
corrections to the asymptotic theory may become important 
even for large coupling parameters (this can be understood also in terms of the
criterion introduced in Sections \ref{subsec:sc_valid_rods} and \ref{sec:spheres}). 
These effects might result in finite corrections for the predicted 
threshold of attraction or scaling exponents, 
such as Eq. (\ref{eq:scaling}), in the limit of large couplings.
However, as we showed, the de-condensation process at small Manning parameters
is captured by the SC free energy and a dominant repulsive interaction 
is obtained, which gives a consistent picture for the whole range 
of Manning parameters. Note also that the threshold of condensation obtained 
for two charged rods (Section \ref{subsec:thre_rods}) coincides 
with the value obtained from Manning theory \cite{Manning97,Manning69}.
Further numerical studies in this regime will be useful to investigate 
the role of higher-order corrections. 
(These aspects have been studied for a single-rod system by means of  
Monte-Carlo simulations \cite{Naji-Netz-unpub}.)

Finally, we emphasize that the thermodynamic behavior of macroionic solutions and
colloidal dispersions may not be derived directly from the SC free energy
presented in Sections \ref{sec:rods} and \ref{sec:spheres}.
As we showed, by decreasing Manning parameter or
increasing the box size, attracting macroions undergo an unbinding transition
to a repulsion-dominated state, 
which occurs continuously for two rods and discontinuously 
for two spheres. These transitions
do not necessarily represent thermodynamic phase transitions, because
in our study, the distance between macroions is
assumed to be fixed, while in realistic situations (as well as in the 
simulations
on charged spheres cited in Section \ref{sec:spheres}), the 
macroion-macroion separation is an annealed degree of
freedom contributing separately to the partition function. 
Also, for fluctuating spheres there will be a logarithmic contribution to the 
entropy for large separations due to the increasing free volume available
for the sphere-sphere distance coordinate.
On the other hand, there has been evidence of a thermodynamic phase
transition of the first order (phase separation) 
from recent numerical simulations on highly-charged spheres
\cite{Gron98,LinsePRL,Lise00,Hribar}.
The systematic study of such phase transitions in the strong-coupling limit 
will be subject of a future investigation.




\begin{acknowledgement}
We are grateful to H. Boroudjerdi, R. Golestanian, 
A.G. Moreira and R. Podgornik for useful discussions,
and to A. Arnold and C. Holm for providing us with their recent
numerical results on the two-rod system. 
We acknowledge financial supports from DFG Schwerpunkt Polyelektrolytes
with defined architecture and DFG German-French Network. 
\end{acknowledgement}




\appendix

\section{Asymptotic analysis of the SC free energy}

\subsection{The two-rod system}
\label{subsec:asymp_rods}

Let us first consider the limit of 
vanishingly small Manning parameter ${\tilde R}\ll 1$. In this case, the
 integral in Eq. (\ref{eq:F1rods}) may be expanded in terms of 
${\tilde R}$, and the rescaled SC free energy per unit length, ${\tilde H}$,
of the rods is subsequently obtained as 
\begin{eqnarray}  
\frac{{\mathcal F}_1}{\tilde H} \approx &-&2 {\tilde R}^2\ln {\tilde D}  
                           +\frac{4{\tilde R}^2}{{\tilde L}^2}
                                \int_{-{\tilde L}/2}^{{\tilde L}/2}
                                \,d{\tilde x}d{\tilde y}\,
                                    [\ln {\tilde r}_1+\ln {\tilde r}_2]
\nonumber\\
&-&2{\tilde R}\ln {\tilde L}^2,
\label{eq:F1rodsRsmallApp}
\end{eqnarray}  
where we have omitted terms that are independent of ${\tilde L}$ and 
${\tilde D}$ as they are irrelevant in our discussion.
The asymptotic expression (\ref{eq:F1rodsRsmallApp}) involves
the rescaled mean energy of the 
system of two-rods and neutralizing counterions (the first two terms)
together with the entropic contribution of the 
counterions (the last term), which has an ideal-gas form. 
(This is seen more clearly using
Eqs. (\ref{eq:FSCF1}) and (\ref{eq:F1rodsRsmallApp}) and restoring 
the actual units that yields
${\mathcal F}_N^{SC}\approx -2\ell_B\tau ^2 H\ln D + 2\xi N \langle \ln r_1 + \ln r_2 \rangle-N\ln L^2$
per $k_BT$ and up to some additive terms.
Note that Manning parameter equals the rescaled radius of rods, 
Eq. (\ref{eq:Rrods}), thus ${\tilde R}\rightarrow 0$ implies 
line charges in the rescaled picture.)
This form of the SC free energy shows that 
at small Manning parameters, 
the counterions are effectively unbound, though still interacting
with the charged rods.
The bare repulsive force between the rods coming from the first term 
scales like $\sim{\tilde D}^{-1}$, but the counterion-induced attraction 
force coming from the second term scales as $\sim {\tilde D}/{\tilde L}^2$, which
becomes vanishingly small as ${\tilde L}\rightarrow \infty$. Therefore, 
the asymptotic free energy (\ref{eq:F1rodsRsmallApp}) 
admits a shallow ${\tilde L}$-dependent minimum as it is seen in 
Figure \ref{fig:freeRsmall}.
By minimizing expression (\ref{eq:F1rodsRsmallApp})
with respect to ${\tilde D}$, the bound-state separation
is approximately found as in Eq. (\ref{eq:D0_txt}) 
for  ${\tilde L}\rightarrow \infty$.

For large ${\tilde R}>{\tilde R}_c$, 
as seen in Figure \ref{fig:freeRlarge}, the free 
energy substantially decreases, when the rods are close to each other. 
Inspection shows that in this situation,
the main contribution to the integral $I$ in Eq. (\ref{eq:F1rods})
comes from the intervening region between the rods
$(-{\tilde D}/2+{\tilde R}<{\tilde x}<{\tilde D}/2-{\tilde R}; 
{\tilde y}\approx 0)$. In fact, this is associated with strong
accumulation of counterions  in this region 
for increasing Manning parameter, as it can be checked directly 
from the counterionic distribution function obtained in the SC
limit \cite{Note3}. The integral $I$, Eq. (\ref{eq:Irods}), 
may be rewritten as
\begin{eqnarray}
  I=\int d{\tilde x}d{\tilde y}  {\tilde \Omega}\,
                               e^{-2{\tilde R} g({\tilde x}, {\tilde y})}, 
\end{eqnarray}
where 
\begin{equation}
  g({\tilde x}, {\tilde y})=[\ln {\tilde r}_1+\ln {\tilde r}_2],
\end{equation} 
and ${\tilde r}_{1,2}$ are defined in Eq. (\ref{eq:r1}). It turns out that 
$({\tilde x}, {\tilde y})=(0,0)$ is the saddle point of $g$, thus for 
sufficiently large ${\tilde R}$, we may use a saddle-point approximation 
to calculate $I$, which gives (up to some irrelevant prefactors)
\begin{eqnarray}
  I\approx e^{-2{\tilde R}\ln {\tilde D}^2}\times {\tilde D}^2\times
                     \int_{-1/2+{\tilde R}/{\tilde D}}^{1/2-{\tilde R}/{\tilde D}} 
                      \,d{\tilde x}\, 
                        e^{8{\tilde R}{\tilde x}^2},
\label{eq:Isaddle_rods}
\end{eqnarray} 
where we have rescaled the coordinates as 
${\tilde x}\rightarrow {\tilde x}/{\tilde D}, {\tilde y}\rightarrow {\tilde y}/{\tilde D}$
(Eq. (\ref{eq:xyrescale})) and assumed that the box size is sufficiently large. 
Indeed, the above approximation is valid only for
sufficiently large ${\tilde R}/{\tilde D}$ ratio, since $g$ is singular at 
$({\tilde x}/{\tilde D}, {\tilde y}/{\tilde D})=(\pm 1/2, 0)$. In other words, 
it remains valid as long as the surface-to-surface separation of rods 
is sufficiently small, {\em i.e.} 
$\delta=({\tilde D}-2{\tilde R})/2{\tilde R}\ll 1$. In this situation,  
the following approximate expression is obtained for the rescaled 
free energy using Eqs. (\ref{eq:Isaddle_rods}) and (\ref{eq:F1rods}),    
\begin{equation} 
  \frac{{\mathcal F}_1}{\tilde H} \approx 6 {\tilde R}^2\ln {\tilde D}
              -2 {\tilde R} f({\tilde D}, {\tilde R}),
 \label{eq:F1largeRrodsApp}
\end{equation} 
where $f\approx \ln ({\tilde D}-2{\tilde R})+{\mathcal O}(\delta)$. 
Now, by minimizing 
expression (\ref{eq:F1largeRrodsApp}) with respect to ${\tilde D}$, 
the equilibrium axial separation is obtained as in 
Eq. (\ref{eq:DlargeR_rods_txt}) for ${\tilde R}\gg 1$.
We emphasize that Eq. (\ref{eq:F1largeRrodsApp}) represents
the approximate form of the free energy around its local minimum, and
is not valid for large separations (or small Manning parameters
close to the attraction threshold ${\tilde R}_c=2/3$). 
The large-distance behavior of the free energy is given by 
Eqs. (\ref{eq:asyF1rods}) and (\ref{eq:asyF1rods_2}).

Restoring the actual units in Eq. (\ref{eq:F1largeRrodsApp}), 
and noting that the rescaled free energy 
${\mathcal F}_1$ is related to the actual SC free energy
(per $k_BT$), ${\mathcal F}_N^{SC}$, 
through Eq. (\ref{eq:FSCF1}), we find 
\begin{equation}
 {\mathcal F}_N^{SC}\approx 6\ell_B\tau^2 H\ln D-N\ln (D-2R).
\label{eq:F1large_rods_actu}
\end{equation}
The first term in this equation formally 
corresponds to the energetic attraction between two rods mediated by
neutralizing counterions that are located between them (lining up 
in $z$ direction with an effective linear charge density of $2\tau$). 
While the second term may be regarded as an (repulsive) entropic 
contribution from counterions. Note that the attractive component
of the resultant force
is independent of the temperature, which only appears as a prefactor 
in the second term. Clearly, for increasing Manning parameter
(or decreasing temperature), the strength of attraction increases \cite{Gron97,Ha}
and saturates to the maximal value given by the first term in 
Eq. (\ref{eq:F1large_rods_actu}).

\subsection{The two-sphere system}
\label{subsec:asymp_sph}

For vanishingly small Manning parameter ${\tilde R}\ll 1$, one may expand 
the integral involved in
the rescaled SC free energy,  Eq. (\ref{eq:F1sph}), which yields  
\begin{eqnarray}
  {\mathcal F}_1 & \approx & 4\frac{{\tilde R}^4}{{\tilde D}}
    -\frac{8{\tilde R}^4}{{\tilde L}^3}
                                \int_{-{\tilde L}/2}^{{\tilde L}/2}
                                \,d^3{\tilde r}\,
                            [\frac{1}{\tilde r}_1+\frac{1}{\tilde r}_2]
     \nonumber\\
    &-&4{\tilde R}^2\ln {\tilde L}^3.
\label{eq:F1sphRsmall}
\end{eqnarray}
This expression is basically similar to the low-Manning-parameter
expansion for the two-rod system, Eq. (\ref{eq:F1rodsRsmallApp}), 
containing the mean energy of the 
two spheres with neutralizing counterions 
(the first two terms) together with the entropic contribution of the 
counterions (the last term) that has an ideal-gas form. 
(This is seen using Eqs. (\ref{eq:FSCF1}) and (\ref{eq:F1sphRsmall}) 
and restoring the actual units that gives
${\mathcal F}_N^{SC}\approx \ell_B Z^2/D-N\ell_B Z q\langle 1/r_1+1/r_2\rangle-N\ln L^3$
in units of $k_BT$.) The bare repulsive force coming from the first term 
scales like $\sim{\tilde D}^{-2}$, but the counterion-induced attraction 
force coming from the second term scales as $\sim {\tilde D}/{\tilde L}^3$, which
becomes vanishingly small as ${\tilde L}\rightarrow \infty$. The asymptotic
free energy (\ref{eq:F1sphRsmall}) has a shallow ${\tilde L}$-dependent minimum as 
seen in Figures \ref{fig:freeS}a-c, the location of which is approximately 
obtained as in Eq. (\ref{eq:D0sph}) in the text.

For large Manning parameter,
as  seen in Figure \ref{fig:freeS}d, the free energy exhibits a local 
minimum at small center-to-center separations, where the main
contribution to the  volume integral in Eq. (\ref{eq:F1sph}) 
comes from the intervening region between spheres
$(-{\tilde D}/2+{\tilde R}<{\tilde x}<{\tilde D}/2-{\tilde R}; 
{\tilde y}\approx 0;{\tilde z}\approx 0)$. In this case,  
a saddle-point approximation similar to the two-rod system 
(Eq. (\ref{eq:Isaddle_rods})) may be performed, which gives the following 
asymptotic expression for the rescaled free energy for sufficiently large 
${\tilde R}$ and small surface-to-surface separation
$\delta=({\tilde D}-2{\tilde R})/2{\tilde R}\ll 1$,
\begin{equation}
  {\mathcal F}_1 = -28\frac{{\tilde R}^4}{\tilde D}
       -4{\tilde R}^2 f({\tilde D}, {\tilde R}),
\label{eq:F1sphRlarge}
\end{equation} 
where 
$f\approx \ln ({\tilde D}-2{\tilde R})+{\mathcal O}(\delta)$. 
The equilibrium separation is subsequently obtained 
by minimizing expression  (\ref{eq:F1sphRlarge}) with respect to ${\tilde D}$
yielding Eq. (\ref{eq:DlargeRsph}) in the text.

Using the free energy expression (\ref{eq:F1sphRlarge}) and restoring 
the actual units, the approximate form of the actual SC free energy, 
Eq. (\ref{eq:FSCF1}), is obtained as
\begin{equation}
   {\mathcal F}_N^{SC}\approx  -7\ell_B\frac{Z^2}{D}-N\ln (D-2R).
\label{eq:F1large_sph_actu}
\end{equation}
The first term in Eq. (\ref{eq:F1large_sph_actu}) formally 
corresponds to the energetic attraction between two spheres mediated by
a neutralizing  counterion (of charge valency $q=2Z$) located between
them. The second terms may be regarded as an (repulsive) entropic 
contribution from counterions. Upon increasing Manning parameter
(or decreasing the temperature), the strength of attraction increases
and saturates to the maximal value given by the first term in 
Eq. (\ref{eq:F1large_sph_actu}), which gives a temperature-independent force
that scales with the center-to-center distance as $\sim 1/D^2$.




\bibliographystyle{}

\end{document}